\documentclass[11pt,a4paper,oneside]{article}

\usepackage{amsmath}%
\usepackage{amsfonts}%
\usepackage{amssymb}%
\usepackage{adjustbox}%
\usepackage{graphicx}%
\usepackage{graphics,color}%
\usepackage[sort]{natbib}%
\usepackage{breqn}%
\usepackage{setspace}
\usepackage{amsthm}
\usepackage{mathtools}
\usepackage[top=1.7in, bottom=1.7in, left=0.9in, right=0.9in]{geometry}
\usepackage{bigints}
\usepackage{epstopdf}
\usepackage{subfigure}
\usepackage{lineno}
\usepackage{setspace}
\usepackage{authblk}
\begin{document}

\title{Bayesian Estimation of the Threshold of a Generalised Pareto Distribution for Heavy-Tailed Observations
}

\author{Cristiano Villa
}

\affil{School of Mathematics, Statistics and Actuarial Science, University of Kent, UK \\
         email: cv88@kent.ac.uk}

\date{}

\maketitle

\begin{abstract}
In this paper, we discuss a method to define prior distributions for the threshold of a generalised Pareto distribution, in particular when its applications are directed to heavy-tailed data. We propose to assign prior probabilities to the order statistics of a given set of observations. In other words, we assume that the threshold coincides to one of the data points. We show two ways of defining a prior: by assigning equal mass to each order statistic, that is a uniform prior, and by considering the \emph{worth} that every order statistic has in representing the true threshold. Both proposed priors represent a scenario of minimal information, and we study their adequacy through simulation exercises and by analysing two applications from insurance and from finance.

\end{abstract}

%--- INTRODUCTION ---------------------------------------------
\section{Introduction}\label{sc_intro}
The purpose of this paper is to outline a novel Bayesian approach to estimate the threshold of a generalised Pareto distribution (GPD) by means of data dependent priors on the order statistics. The statistical model for the overall sample is a mixture model with two main components: a model for the non-extreme data below a certain threshold, also labelled as the \emph{bulk data}, and the GPD to model the extreme values above the threshold. The component for the bulk data do not represent our main concern; therefore, we will be using a finite mixture of densities where the components will somehow reflect the nature of the phenomenon of interest \citep{Donasc:2011}. In particular, a mixture of gamma densities if we are interested in positive data (e.g. insurance losses, river floods, rainfall), and a mixture of normal densities for data that can take both positive and negative values (e.g. financial returns). The second component of the overall model is a GPD where the threshold parameter $\theta$, conceptually separating non-extreme from extreme observations, has an assigned uncertainty represented by a prior probability distribution. The details of the overall model will be discussed in Section \ref{sc_model}. \\

The idea behind extreme value theory is that the main interest is in the tail (or tails) of a distribution. In areas such as finance, insurance, environmental sciences and engineering, the focus is often on observations that present a clear difference in value from the bulk data. Due to this extremal nature of some observations, a distribution that model the whole data would not be appropriate as the majority of observations used to estimate the parameters are non-extreme. It is then necessary to use an appropriate procedure that, whilst still allowing for a reasonable inference of the bulk data, permits a precise estimate of the main characteristics of the tail observations. Depending on the area of application, justifications of the adoption of extreme value distributions can be found, for example, in \cite{Fabozzi:2010} for finance, \cite{DonEm:2010} for insurance and actuarial science; or, to cover a wider range of applications, including environmental sciences and engineering, refer to \cite{Coles:2001} and \cite{Smith:1984}.

To set the scene, suppose we have observed the sample $x=(x_1,\ldots,x_n)$ from a model with distribution function $F(x)$. Under some specific conditions \citep{Pick:1975}, the distribution of $x$ above a certain value $\theta$ can be approximated by a GPD with distribution function

\begin{equation}\label{eq_intro_1}
G(x|\xi,\sigma,\theta) = 
	\begin{cases}
	1-\left\{1+\dfrac{\xi(x-\theta)}{\sigma}\right\}^{-1/\xi} & \text{if } \xi\neq0 \\
	1-\exp\left\{-\dfrac{x-\theta}{\sigma}\right\} & \text{if } \xi=0,
					\end{cases}
\end{equation}

\noindent
where $\sigma>0$ is the scale parameter and $\xi$ is the shape parameter. The support for \eqref{eq_intro_1} is $x\geq \theta$ for $\xi\geq0$ and $\theta\leq x\leq -\sigma/\xi$ for $\xi<0$. Although later the case $\xi<0$ will be briefly mentioned, the main focus of the paper is for $\xi>0$, where the GPD presents a heavy-tailed behaviour. The assessment of the threshold $\theta$ is critical. In fact, if its value is not large enough, the resulting model would be incorrect as the asymptotic tail approximation discussed in \cite{Pick:1975} is no longer valid. On the other hand, if the value $\theta$ is too high, then the number of observations above it would not be sufficient to have reasonably precise estimates of the parameters $\xi$ and $\sigma$. \\

The idea of using order statistics to identify the threshold of a GPD is not new. In the context of a Bayesian predictive approach, \cite{Dezea:2001} assigns a discrete prior to the number of upper order statistics, that is to the number of observations that could be classified as excedances. What we propose here has a different flavour, and it assumes that the threshold corresponds to one of the observed data points.
The detailed motivations for a discrete prior for the threshold of a GPD will be given in Section \ref{sc_model}, when the overall statistical model is introduced. In short, if the data are modelled by a mixture of two components, one for the data below the threshold and one for the data above the threshold, then the assumption of having the threshold coinciding with one of the order statistics is sensible for the following two reasons: there is no evidence about the threshold value between any two consecutive order statistics, and the contribution of each excedance to the GPD likelihood is maximised when the threshold lies on an order statistic.
We propose two different criteria to define a prior distribution on the order statistics. The first one assigns equal mass to each order statistics, and the second one assigns a mass which depends on the \emph{worth} that each order statistics, as the potential threshold, has as being part of the model \citep{Villa:Walker:2014}. Although the proposed priors tend to yield posterior distributions with similar frequentist properties, we discuss some situations and reasons where either one or the other have to be preferred. In addition, although for different reasons, the proposed prior distributions can be categorized as \emph{objective}, as defined in \cite{Berger:2006}, and are suitable to be employed in scenarios of minimal prior information.\\

The outline of the paper is as follows. In Section \ref{sc_model} we discuss the details of the mixture model for the whole data set and the prior distributions for the parameters. For the priors, we place the main focus on the prior distributions we propose for the threshold of the GPD. We then conduct simulation studies in Section \ref{sc_simul} by first illustrating how the proposed priors for $\theta$ apply to a single independent and identically distributed sample, and then by analysing the frequentist performances of the respective induced posterior distributions. Section \ref{sc_real} is dedicated to applying the defined model and the proposed prior distributions for the threshold to real data examples; in particular, we analyse the well known data set of the Danish fire loss, and the daily increments of the NASDAQ-100 index over a period of more than seventeen years. Finally, the concluding discussion and remarks are presented in Section \ref{sc_disc}.

%--- MODELS AND PRIOR -----------------------------------------
\section{The model and the priors}\label{sc_model}
\subsection{The mixture model}
The model considered in this paper has two components: a finite mixture of parametric distributions for the data below the threshold $\theta$, and a GPD for the data above the threshold. If we represent the distribution function for the bulk data by $H(\cdot|\gamma)$, the distribution function for the whole set of observations is given by

\begin{equation}\label{eq_overallmixture}
F(x|\gamma,\xi,\sigma,\theta) = 
	\begin{cases}
	H(x|\gamma) & x<\theta \\
	H(\theta|\gamma)+[1-H(\theta|\gamma)]G(x|\xi,\sigma,\theta) & x\geq \theta,
					\end{cases}
\end{equation}

\noindent
where $G(x|\xi,\sigma,\theta)$ is the distribution defined in \eqref{eq_intro_1}, and $\gamma$ represents the parameters of the mixture for the bulk data. More in general, the mixture model in \eqref{eq_overallmixture} can be categorised on the basis of the nature of $H(\cdot|\gamma)$: parametric bulk model, semiparametric bulk model and nonparametric bulk model \citep{Scarr:2012}.

An example of the first type consider a gamma distribution for the bulk data \citep{Behr:2004}. Of course, other parametric distribution can be considered, such as the normal, the lognormal or the Weibull, so to reflect a different nature of the data. The main drawback of parametric bulk models is the lack of flexibility, resulting in a difficult identification of the threshold, except when the processes generating the bulk data and the extreme data are well discernible \citep{Scarr:2012,Behr:2004}. To overcome this difficulty, semiparametric bulk models have been proposed. \cite{Castel:2011} propose a spliced model for the bulk data, while \cite{Donasc:2011} discuss a finite mixture of gamma densities. Examples of a nonparametric approach for the bulk data can be found in \cite{Tancredi:2006}, \cite{Macetal:2011} and \cite{Pati:2015}. All the above references concern Bayesian approaches to deal with the GPD. Recent publications discussing different approaches worthwhile to be mentioned are \cite{NorCol:2014} and \cite{WadTaw:2012}, among others.

The focus of this paper is on the determination of the threshold $\theta$, and we represent the bulk data with a finite mixture of distributions \citep{Donasc:2011}. As discussed by the authors, the approach allows for appropriate adaptation and, therefore, flexibility of the overall mixture model. For a more detailed discussion of this specific type of models for the bulk data and, in general, about semiparametric models, we refer to \cite{Scarr:2012} and \cite{Donasc:2011}.

\subsection{The prior for the threshold}\label{sc_thetaprior}
The main contribution of this work is in the prior for the threshold $\theta$ of a GPD. In fact, we propose to assign a prior probability to the observed order statistics by assuming that $\theta=x^{(k)}$, where in general $k$ can take any value in $\{1,\ldots,n\}$. Therefore, the nature of the proposed priors is of a discrete \emph{data dependent} prior. Before outlining how a prior on the order statistics can be defined, we need to fully motivate the choice of a distribution which support is limited to the order statistics only.

The density of model \eqref{eq_overallmixture2} has the form
\begin{equation}\label{eq_overallmixture2}
f(x|\gamma,\xi,\sigma,\theta) = 
	\begin{cases}
	h(x|\gamma) & x<\theta \\
	[1-H(\theta|\gamma)]g(x|\xi,\sigma,\theta) & x\geq \theta,
					\end{cases}
\end{equation}
where $h(x|\gamma)$ is the density of the bulk data mixture, and $g(x|\xi,\sigma,\theta)$ the density of a GPD. Note that, being \eqref{eq_overallmixture2} a mixture model with two components, it can also be represented as
\begin{equation}\label{eq_overallmixture3}
f(x|\gamma,\xi,\sigma,\theta) = \omega f_1(x|\theta) + (1-\omega) f_2(x|\theta),
\end{equation}
where $\omega=P(X<\theta)$, $f_1(x|\theta)=h(x|\gamma)/H(\theta|\gamma)\cdot1_{(-\infty,\theta)}(x)$, and $f_2(x|\theta)=g(x|\xi,\sigma,\theta)\cdot1_{[\theta,\infty)}(x)$. As in this work we consider quantities that can take positive values only, we will have $f_1(x|\theta)=h(x|\gamma)/H(\theta|\gamma)\cdot1_{(0,\infty)}(x)$. If we observe sample $(x_1,\ldots,x_n)$, which results in the order statistics $(x^{(1)},\ldots,x^{(n)})$, the likelihood function of model $\eqref{eq_overallmixture2}$ (or, equivalently, model \eqref{eq_overallmixture3}) is given by
\begin{equation}\label{eq_like1}
L(\gamma,\xi,\sigma,x^{(k)}|x) = \prod_{j<k}h\left(x^{(j)}|\gamma\right) \times \prod_{j\geq k}\left[1-H(x^{(k)}|\gamma)\right]g\left(x^{(j)}|\xi,\sigma,x^{(k)}\right),
\end{equation}
where we have assumed that the threshold of the GPD satisfies $x^{(k-1)}<\theta\leq x^{(k)}$, with $k=2,\ldots,n$. Note that, although the impact of the observations below the threshold  on the estimates of the GPD parameters is in general not prominent \citep{Scarr:2012}, we still deem appropriate to consider it, and it is therefore included in the likelihood. For reasons due to practicality and identifiability of the model \citep{Castel:2011}, we assume that at least one observation contributes to the likelihood of the bulk component of the overall model. As mentioned in Section \ref{sc_intro}, from \eqref{eq_like1}, we note that observations $x^{(1)},\ldots,x^{(k-1)}$ contribute to the bulk part of the model $h(x|\gamma)$, while observations $x^{(k)},\ldots,x^{(n)}$ contribute to the GPD part of the overall mixture model. As such, there is no information for any $\theta$ within the interval $(x^{(k-1)},x^{(k)})$, and the choice to assume that the threshold coincides to one of the order statistics is sensible. An additional argument, though connected to the above one, can be made by considering the following characteristics of the density of the GPD, which has the form
\begin{equation}\label{eq_gpddensity1}
g(x|\xi,\sigma,\theta) = \sigma^{-1}\left\{1+\frac{\xi}{\sigma}(x-\theta)\right\}^{-(1+\xi)/\xi}, \qquad \xi\neq0.
\end{equation}
At least for the case of interest in this paper, that is $\xi>0$, the density \eqref{eq_gpddensity1} is decreasing. Therefore, if $x^{(k-1)}<\theta\leq x^{(k)}$, the choice of $\theta=x^{(k)}$ is optimal in the sense that the contribution of the excesses $(x^{(k)}-\theta,x^{(k+1)}-\theta,\ldots,x^{(n)}-\theta)$ to the GPD part of the likelihood is maximised. That is, any other choice of $\theta<x^{(k)}$ would yield a smaller contribution of the excesses to the GPD likelihood. We should also not forget that the threshold of the GPD is an artificial parameter \citep{Donasc:2011}, defining at what point of the support it is safe to assume tail approximation, and its determination within a given interval $(x^{(k-1)},x^{(k)}]$ is not driven by any information in the sample beyond the interval boundaries themselves. As such, in a mixture model set up as the one considered in this work, the choice of having $\theta$ equal to an order statistics is appropriate. \\

We propose two discrete priors for the threshold, both of which can be seen as the result of a choice under minimal prior information. The first proposed prior distribution is a discrete uniform prior. It is assumed that the threshold coincides with one of the observed order statistics; in other words, the parameter space for the threshold of the GPD is $\Theta=\{x^{(2)},\ldots,x^{(n)}\}$, and the prior can be written as $\pi(k)$ or $\pi(x^{(k)})$, with $k=2,\ldots,n$. In this work we will be using the latter notation, leaving $\theta$ to represent the true (or theoretical) threshold value. From a practical point of view, the choice of a finite uniform prior to represent prior minimal information is obvious and, in some sense, intuitive. The prior has computational advantages and it is easy to be implemented. From \cite{Castel:2011}, where a continuous uniform prior is proposed, we see that the uniform should be defined over the interval $(x^{(m+1)},x^{(n-2)})$, where $m$ is the number of parameters of the model for the bulk component (i.e. the dimension of $\gamma$). In this case, the overall prior will be
$$\pi(\gamma,\xi,\sigma,x^{(k)}) \propto \pi(\xi,\sigma)\pi(\gamma),$$
where the parameters ($\xi$, $\sigma$) and $\gamma$ are in general assumed to be independent a priori. The assumption of considering the parameters of the model for the bulk data independent from the parameters of the GPD is sensible. In fact, the general idea is that we have a set of observations that have been generated by two different processes. Therefore, the information of the first process that impacts the second process (and vice-versa) can be assumed to be on the threshold only \citep{Scarr:2012}. In addition, parametric mixture models, such as in \cite{Behr:2004}, \cite{MenLop:2004} \cite{CarBen:2009a}, have been criticised as they do not take into consideration the dependence between the threshold and the scale $\sigma$ of the GPD \citep{Scarr:2012}. However, since this dependence occurs for $\xi<0$, what discussed in this paper is not subject to the above criticism.

The second prior we propose is based on concepts of loss in information, therefore identified as the prior based on losses (or as the KL prior, for reasons which will become clear below, where KL stands for Kullback--Leibler divergence). In this case the prior for the threshold depends on the parameters of the GPD ($\xi$ and $\sigma$), and the overall prior has the form
$$\pi(\gamma,\xi,\sigma,x^{(k)}) = \pi(x^{(k)}|\xi,\sigma)\pi(\xi,\sigma)\pi(\gamma).$$
The idea used to obtain the prior $\pi(x^{(k)}|\xi,\sigma)$ is derived from \cite{Villa:Walker:2014} and it is as follows. Let us assume to have observed data $x=(x_1,\ldots,x_n)$. Given the order statistics $x^{(1)},\ldots,x^{(n)}$, we also assume that the true value of the threshold is $\theta=x^{(k)}$, with $k=2,\ldots,n$. We consider the threshold to be $\theta>x^{(1)}$, so that there will be at least one observation from the bulk distribution. Should the inferential process suggest $\theta=x^{(n)}$, it would then be practical (and sensible) to assume that a GPD is not necessary and that the model for the bulk data is a better choice.

It is important to remark that the component $h(\cdot|\gamma)$ of the model is not considered in the construction of the prior for the threshold. The main reason being that, with the type of problems considered in this paper, the focus is on the tail of the model, i.e. on the extreme values. In addition, the mixture model approach here discussed is thought in a way that the mixture distribution for the bulk data is included for convenience only and little consideration is given to its actual fitting to the data. As such, in order to use the prior information for $\theta$ in a relatively sharp way, it seems more appropriate to use the information of the order statistics above the threshold only, leaving the information coming from the bulk data to contribute by means of the likelihood function.

The prior mass to be put on $x^{(k)}$ is derived by considering what is lost if the model $g(x|\xi,\sigma,x^{(k)})$ is removed and it is the true one, where $g(\cdot|\xi,\sigma,x^{(k)})$ is the density of a GPD with threshold $x^{(k)}$, shape parameter $\xi$ and scale parameter $\sigma$. In other words, the approach associates a \emph{worth} to each parameter value which, in this particular circumstance, is derived from the fact of having observed a particular value of $x$. The \emph{worth} is measured by applying a result in \cite{Berk:1966} which states that, if a model is misspecified, i.e. if $x^{(k)}$ is removed and it is the true threshold, then the posterior distribution asymptotically accumulates at the order statistics $x^{(k^\prime)}$ such that the Kullback--Leibler divergence \citep{Kull:1951} $D_{KL}(g(\cdot|\xi,\sigma,x^{(k)})\|g(\cdot|\xi,\sigma,x^{(k^\prime)}))$ is minimised. That is, if the true model is removed, the estimation process will asymptotically indicate as the correct model the nearest one, in terms of the Kullback--Leibler divergence; viz., the model which is the most similar to the true one \citep{BerSmi:1994}.

To link the \emph{worth} of each order statistics to the prior probability we use the \emph{self-information} loss function. This particular type of loss function assigns a loss to a probability statement and, say we have defined prior $\pi(x^{(k)}|\xi,\sigma)$, its form is $-\log\pi(x^{(k)}|\xi,\sigma)$. More information about the self-information loss function can be found, for example, in \cite{Merhav:Feder:1998}. To formally derive the prior for the threshold we can proceed in terms of utilities, instead of losses; this approach allows for a clearer exposition and does not impact the logic behind the prior derivation. Let us then write utility $u_1(x^{(k)})=\log\pi(x^{(k)})$ where, to simplify the notation, we have dropped parameters $\xi$ and $\sigma$. We then let the minimum divergence from $x^{(k)}$ to be represented by utility $u_2(x^{(k)})$. We want $u_1(x^{(k)})$ and $u_2(x^{(k)})$ to be matching utilities functions, as they measure the same utility in $x^{(k)}$; though as it stands $-\infty<u_1\leq0$ and $0\leq u_2<\infty$, and we want $u_1=-\infty$ when $u_2=0$. The scales are matched by taking exponential transformation; so $\exp(u_1)$ and $\exp(u_2)$ are on the same scale. Hence, we have
\begin{equation}\label{eq_prior_1}
e^{u_1(x^{(k)})} = \pi(x^{(k)})\propto e^{w[u_2(x^{(k)})]},
\end{equation}
where
\begin{equation}\label{eq_prior_2}
w(u) = \log(e^u-1).
\end{equation}
By setting the functional form of $w$ in \eqref{eq_prior_1} as it is defined in \eqref{eq_prior_2}, we derive the objective prior distribution for the order statistics
\begin{equation}\label{eq_prior_3}
\pi(x^{(k)}|\xi,\sigma) \propto \exp\left\{ \min_{k^\prime\neq k} D_{KL}\Big(g(\cdot|\xi,\sigma,x^{(k)})\|g(\cdot|\xi,\sigma,x^{(k^\prime)})\Big) \right\}-1, \qquad k,k^\prime=2,\ldots,n.
\end{equation}

To identify the minimum Kullback--Leibler divergence in \eqref{eq_prior_3}, we first consider

\begin{eqnarray}\label{eq_prior_4}
D_{KL}(g(x|\xi,\sigma,x^{(k)})\|g(x|\xi,\sigma,x^{(k+c)})) &=& \int_{x^{(k)}}^\infty g(x|\xi,\sigma,x^{(k)})\nonumber\\
&&\log\left\{ \frac{g(x|\xi,\sigma,x^{(k)})}{g(x|\xi,\sigma,x^{(k+c)})} \right\}\;dx \nonumber \\
&=& -\frac{1+\xi}{\xi}\left\{ \mathbb{E}\left[ \log\left(1+\frac{\xi}{\sigma}\Big(x-x^{(k)}\Big)\right) \right] \right. \nonumber\\
&& \left. - \mathbb{E}\left[ \log\left(1+\frac{\xi}{\sigma}\Big(x-x^{(k+c)}\Big)\right) \right]  \right\}
\end{eqnarray}
where $c\neq0$ and the expectations are taken with respect to the density $g(x|\xi,\sigma,x^{(k)})$. As \eqref{eq_prior_4} is decreasing in $c$, the nearest GPD to $g(x|\xi,\sigma,x^{(k)})$ is either $g(x|\xi,\sigma,x^{(k-1)})$ or $g(x|\xi,\sigma,x^{(k+1)})$. However, given that $g(x|\xi,\sigma,x^{(k+1)})$ is zero for $x\in(x^{(k)},x^{(k+1)})$, resulting in an infinite divergence, the prior is

\begin{equation}\label{eq_prior_5}
\pi(x^{(k)}|\xi,\sigma) \propto \exp\left\{D_{KL}\Big(g\big(x|\xi,\sigma,x^{(k)}\big)\|g\big(x|\xi,\sigma,x^{(k-1)}\big)\Big)\right\}-1.
\end{equation}

The behaviour of the prior \eqref{eq_prior_5} is obvious in the ideal case where the bulk and the extreme data have been generated by two clearly distinct processes. In this scenario there would be a large ``jump'' separating the two sets of data. Prior \eqref{eq_prior_5} will then put the highest mass on the most left order statistics of the extreme set of data, as its nearest model is relatively far. This value would then represent the best candidate of being the threshold separating the extreme from the bulk data, given the information coming from the observations and the choice of the model. In most realistic scenarios observed data would most likely not display an abrupt ``jump'' between the bulk and the extreme components; rather, a smooth transition has to be expected. Sections \ref{sc_simul} and \ref{sc_real} present both simulated and real data scenarios, where it is possible to have a feeling of the shape of the prior based on losses, and how its performances can be compared with the ones of the uniform prior.

In considering the qualitative behaviour of the prior distribution based on losses, we also need to take into account the case where there may be two or more observations with the same value. Although it is possible, and perhaps advisable, to assume that the data are different from each other as this may lead to conceptual issues in the definition of the posterior distribution \citep{FerSteel:1998}, it is easy to see how the proposed prior would behave in this scenario. If we have two (or more) order statistics with the same value, say $x^{(j)}=x^{(j+1)}$, then, by the way the prior is constructed, the mass on $x^{(j+1)}$ would be zero, but the mass on $x^{(j)}$ would be strictly positive, provided $x^{(j)}\neq x^{(j-1)}$. As such, the prior based on losses maintains the idea of assigning mass on the basis on how ``extreme'' a value is, even when there are repeated observations. \\

Given that the prior distributions proposed are data dependent, it is appropriate to briefly discuss the implications of such a choice.

A definition of data dependent prior can be found in \cite{Wass:2000}, who identifies it as a measurable mapping from the data space to the set of priors. In other words, a distribution that depends on the data obtained through avert use of the observations. The above can be accomplished in different ways (and at different levels of depth), but probably the most common type of data dependent priors are the \emph{data-analytic} priors, where the data is used to determine the hyperparameter(s) of the prior distribution. Examples can be found in \cite{Morris:1983}, \cite{Berger:1985}, \cite{Carlin:1990} and \cite{Czado:2005}. Data-analytic priors can also be used to choose the base measure and the precision of a Dirichlet Process in Bayesian nonparametric \citep{Macea:1998,Macaul:2006}. Finally, \cite{Wass:2000} and \cite{Rafte:1996} discuss data-analytic priors for finite mixtures of normal densities.

Although data dependent priors are used in practical situations, criticisms have been raised. Possibly, the most important concerns are that the data is used twice, for the prior and for the likelihood, and that Bayes theorem can only be approximated. An interesting discussion about the first objection can be found in \cite{Gelman:2014}, for example; while the second objection is discussed, for example, in \cite{DeeLin:1981}. We do not present here a detailed discussion on how the above objections can be rebutted or overcome; such a discussion can be found, for example, in the work of \cite{Darn:2011} and the reference therein. Obviously, using the order statistics to determine the parameter space of the threshold categorises our priors under Wasserman's definition of a data dependent prior. In the case of the uniform prior, the information drawn from the data is limited to the possible location of the threshold and, as discussed above, the choice is sensible as it yields optimal contribution of the excesses to the likelihood. For the prior based on the Kullback--Leibler divergence, the information drawn from the data goes beyond the possible location of the threshold, as it considers the similarity (or diversity) between consecutive models. However, as will be shown in Section \ref{sc_simul}, the frequentist performances of this prior are virtually the same as the uniform, showing that the extra information used does not add tangible performance benefits.

To conclude, we deem appropriate to point out that information from the data (besides in the likelihood function) has been always considered in the inferential process for the threshold. This is obvious when we consider graphical approaches \citep{Coles:2001}, where data are plotted to determine a possible location of the threshold. When it comes to Bayesian analysis, the proposed priors in the literature which claim to carry minimal information, draw
some of this information from the data. For example, the continuous uniform prior proposed in \cite{Castel:2011} has a parameter space bound by order statistics. The normal prior proposed by \cite{Behr:2004}, and claimed to be set up in a noninformative fashion by \cite{Donasc:2011}, has to be centered on the $90\%$ data quantile in order to avoid identifiability issues when the sample size is not sufficiently large.

\subsection{The priors for $(\xi,\sigma)$}\label{sc_priorxisigma}
The choice of the prior distribution for the parameters $\xi$ and $\sigma$ of the GPD is straightforward. In a noninformative context, as it is the flavour of this paper, the choice is on the Jeffreys' independent prior defined in \cite{Castel:2007} as
\begin{equation}\label{eq_jeff_1}
\pi(\xi,\sigma) \propto \sigma^{-1}(1+\xi)^{-1}(1+2\xi)^{-1/2},
\end{equation}
which is defined for $\xi>-0.5$ and $\sigma>0$. As shown by \cite{Castel:2007}, the prior \eqref{eq_jeff_1} yields to the proper posterior $\pi(\xi,\sigma|x)$ for a sample size of $n\geq1$. On the other hand, if suitable prior information about $\xi$ and $\sigma$ is available (and it is practical/desirable to be exploited), then appropriate prior distributions can be elicited. However, as this case lies outside the scope of this work, it will not be discussed any further.

\subsection{The prior for $\gamma$}\label{sc_priorgamma}
$\gamma$ is a vector which elements are the parameters of the mixture $h(\cdot|\gamma)$ for the bulk data. Thus the prior to be assigned to $\gamma$ depends on the components of the mixture. As already mentioned, the focus of this work is mainly in the prior for the threshold $\theta$; we then restrict our illustrations to the common case of positive data only and we will adopt a finite mixture of gamma densities to represent the bulk data \citep{Wiper:2001}
\begin{equation*}\label{eq_gammamix1}
h(x|\gamma) = \sum_{j=1}^r \omega_j f_j(x|a_j,b_j).
\end{equation*}
We have $\gamma=(\omega_1,\ldots,\omega_r,a_1,\ldots,a_r,b_1,\ldots,b_r)$, where $(\omega_1,\ldots,\omega_r)$ denote the weights of the mixture, with $\sum\omega_j=1$, $f_j(\cdot|a_j,b_j)$ is a gamma density with shape parameter $a_j$ and rate parameter $b_j$. To address the identifiability issue intrinsic to mixture models \citep{DieRob:1994} the gamma density can be reparametrised as
\begin{equation}\label{eq_gammarep}
f_j(x|\alpha_j,\beta_j) = \frac{\left(\beta_j/\alpha_j\right)^{\beta_j}}{\Gamma(\beta_j)} x^{\beta_j-1} e^{-x\beta_j/\alpha_j}. \qquad j=1,\ldots,r,
\end{equation}
so we can impose the constraint $0<\alpha_1<\cdots<\alpha_r$ on the parameter space for the $\alpha$'s, as they represent the means of the gamma densities. $\beta_1,\ldots,\beta_r$ will represent the shape parameters for the $r$ gamma densities.

With the parametrisation in \eqref{eq_gammarep} we assign an inverse gamma prior to each mean $\alpha$ and a gamma prior to each shape parameter $\beta$. Although the above priors are not selected through an objective method, they will represent minimal prior information in the form of large variance.

Finally, for the weights $\omega_1,\ldots,\omega_r$ we chose a symmetric Dirichlet prior distribution with all the parameters equal to one: $\pi(\omega_1,\ldots,\omega_r)\sim Dir(1,\ldots,1)$. This choice as well represents minimal prior information, and $\pi(\omega_1,\ldots,\omega_r)\propto1$.

%--- SIMULATIONS ----------------------------------------------
\section{Analysis of the posterior distribution for the threshold}\label{sc_simul}
To analyse and compare the proposed discrete priors for the threshold of the GPD we perform two types of simulations. In the first simulation we detail the inferential procedure for all the parameters of the mixture on the basis of a random sample from a known model. The second part consists in a simulation study that aims to assess the frequentist performances of the posterior distributions induced by the proposed priors. This is done by repeatedly sample from mixture models that differ in the GPD component only (i.e. threshold, shape and scale parameters) and observe the coverage and the means square errors of the posterior distributions for the threshold. Given the minimal informative nature of the paper, the analysis of the frequentist properties is a suitable way to compare the two proposed priors and assess their effectiveness.

The posterior for the parameters of the mixture model in \eqref{eq_overallmixture2} is given by
\begin{equation*}\label{eq_posterior}
\pi(\gamma,\xi,\sigma,x^{(k)}|x) \propto L(\gamma,\xi,\sigma,x^{(k)}|x)\times\pi(x^{(k)})\pi(\xi,\sigma)\pi(\gamma),
\end{equation*}
where $L(\gamma,\xi,\sigma,x^{(k)}|x)$ is the likelihood function specified in \eqref{eq_like1}. The prior distribution $\pi(x^{(k)})$, in our illustrations, would be one of the proposed discrete priors; that is, either the uniform prior or the prior based on losses. As the marginal posterior distributions of the parameters are analytically intractable, Monte Carlo methods are necessary to sample from these distributions.

\subsection{Simulation from a single i.i.d. sample}
To illustrate in detail the entire inferential procedure, we have sampled $n=1,000$ observations from a mixture model as in \eqref{eq_overallmixture2}. The bulk data component is a mixture of two gamma densities with shape parameters $a_1=4$ and $a_2=8$, and rate parameters $b_1=2$ and $b_2=1$. The weights of the gamma densities are, respectively, $\omega_1=2/3$ and $\omega_2=1/3$. The extreme data component is a GPD with shape parameter $\xi=0.4$ and scale parameter $\sigma=2$, and the threshold has been put at the $90\%$ data quantile, with $\theta=9$.

\begin{figure}[hbtp]
     \begin{center}
        \subfigure{%
           \label{Prior_1}
           \includegraphics[scale=0.33]{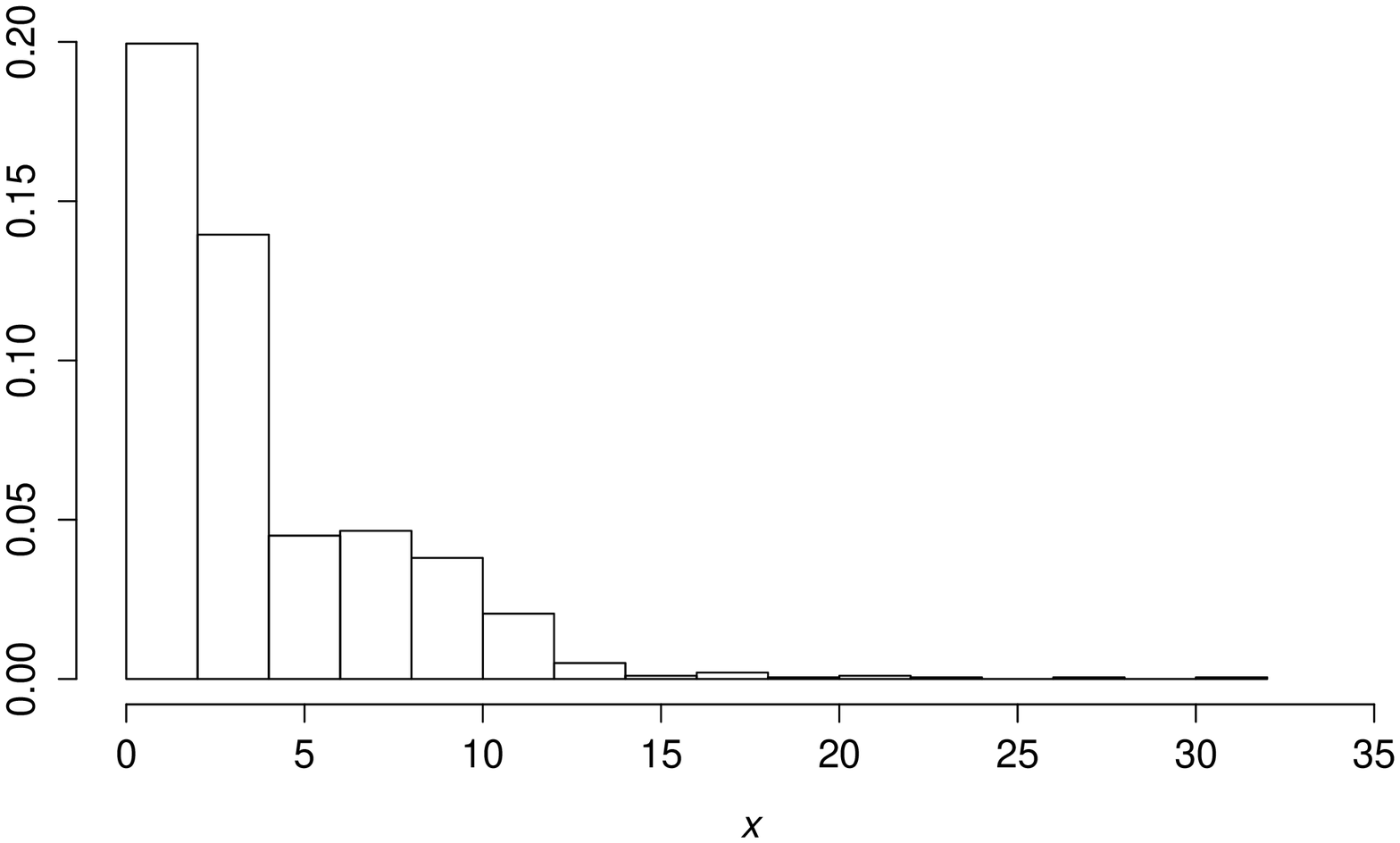} 
        } %
        \subfigure{%
           \label{Prior_2}
           \includegraphics[scale=0.33]{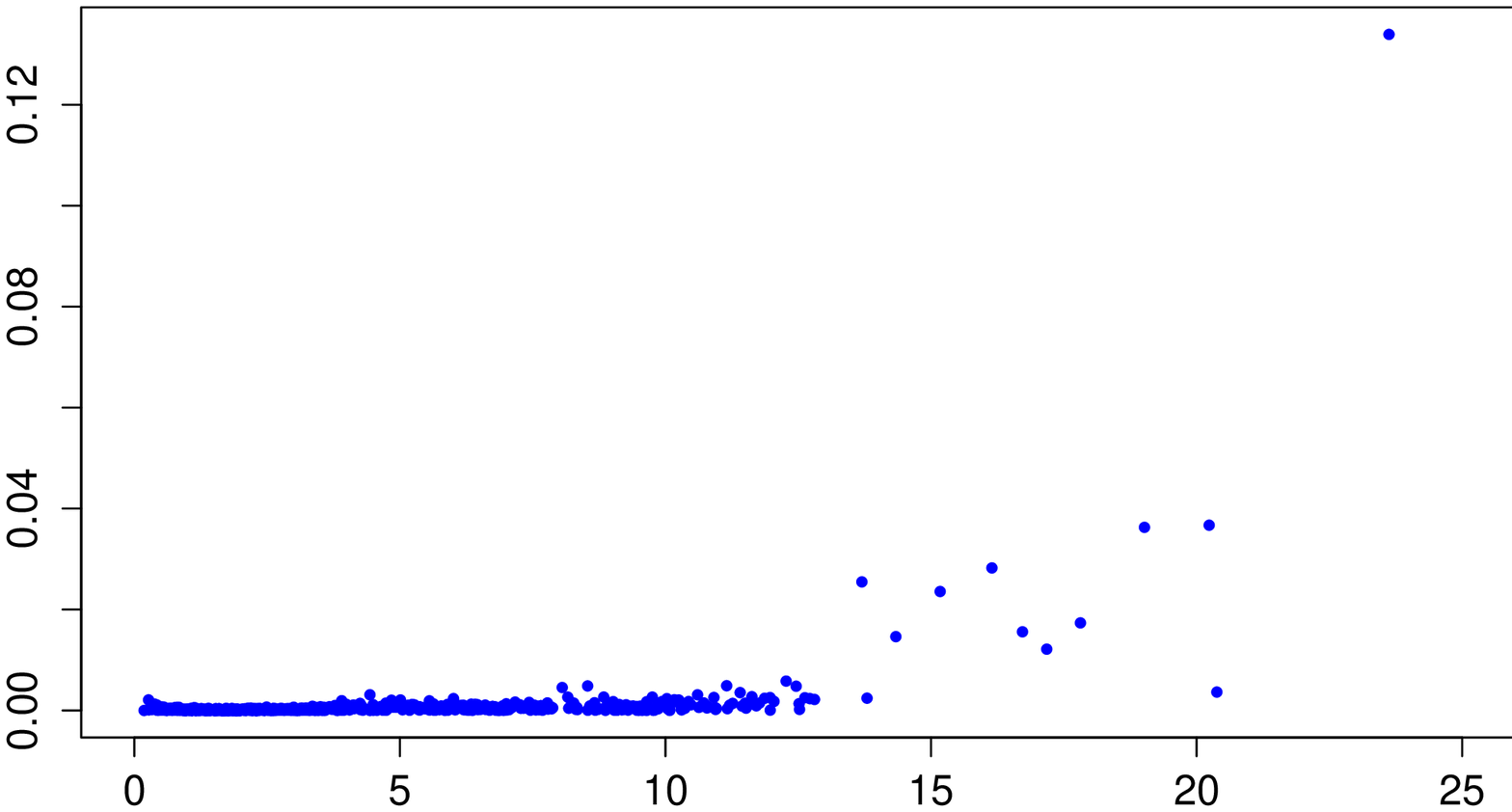} 
        } %  
    \end{center}
    \caption{%
        Histogram of the sample (left) and plot of the discrete prior for $x^{(k)}$ based on losses. The bulk data model is a mixture of two gamma densities, $G_1(4,2)$ and $G_2(8,1)$, with weights $\omega_1=2/3$ and $\omega_2=1/3$, while the extreme data is modelled by a GPD with parameters $\xi=0.4$, $\sigma=2$ and threshold $\theta=9$.
     }%
   \label{fig:simulation_histandprior}
\end{figure}

Figure \ref{fig:simulation_histandprior} shows the histogram of the sample (left graph) and the prior probabilities on the order statistics (right graph) representing the prior for the threshold based on losses. From the histogram we see that there is a smooth transition between the bulk data and the extreme data. The behaviour of the prior for $x^{(k)}$, which we recall being based on the Kullback--Leibler divergence between GPD densities with thresholds on adjacent order statistics, reflects the level of ``extremeness'' of the data: almost uniform for the lower part of the data space, with mass that is assigned increasingly on the order statistics when these become extreme. As discussed in Section \ref{sc_thetaprior}, the behavior of the prior for the threshold as shown in Figure \ref{fig:simulation_histandprior} is sensible as more extreme observations are more likely to represent a suitable threshold.

To estimate the number of components of the mixture for the bulk data ($r$) one could proceed as suggested in \cite{Donasc:2011}, where models with different values of $r$ are estimated and suitable indexes, such as the deviance information criterion (DIC) and the Bayesian information criterion (BIC), are computed to choose the ``best'' model on the basis of the observed sample. Alternatively one could consider a hierarchical structure and assign a prior to $r$ to represent the uncertainty on its true value; for this approach see, for example, \cite{Menger:2011}. We have already mentioned that the focus of this work is on the prior for $x^{(k)}$; therefore, we will not further investigate this matter, and we simply show that the posterior distributions for the weights are different from zero only for $r=2$.

For the parameters of the mixture, as discussed in Section \ref{sc_priorgamma}, we use inverse gamma priors on the means of the gamma densities, and gamma priors on the shape parameters. Given that we want prior distributions that somehow represent weak prior information, these distributions will have large variances. In detail, we have a gamma with parameters 6 and 0.5 for each shape $\alpha_j$, and an inverse gamma with parameters 2.1 and 5.5 for the each mean $\beta_j$, for $j=1,\ldots,r$. In addition, the priors have mean equal to the average of the corresponding true values. For the weights (Section \ref{sc_priorgamma}) we choose a Dirichlet distribution with all parameters equal to one, corresponding to a noninformative scenario. The estimation of $x^{(k)}$ has been performed by considering, for the same sample, both the uniform prior and the prior bases on losses in \eqref{eq_prior_5}.

The Monte Carlo procedure consists of 20,000 iterations with 10,000 iterations as burn-in period. Convergence of the posterior has been assessed by several means, including monitoring the chains, running means and computing the Gelman and Rubin's convergence diagnostics \citep{GelRub:1992}.

First, considering $r=3$, we have seen that the value of $\omega_3$ converged to zero almost immediately under each prior on $x^{(k)}$, which makes us conclude that the model with $r=2$ is the appropriate one. Table \ref{tab:simulation} shows the statistics of the marginal posterior distributions of all the parameters of the mixture model, namely the parameters of the mixture component, and the parameters of the GPD, for $r=2$. These statistics have been computed for both the priors for the threshold.
\begin{table}
\centering
\tabcolsep=0.11cm
\begin{tabular}{|c|c|ccc|ccc|}
\hline 
 &  & \multicolumn{3}{c|}{KL Prior} & \multicolumn{3}{c|}{Uniform Prior} \\ 
\hline 
 & True Value & Mean & Median & $95\%$ C.I. & Mean & Median & $95\%$ C.I. \\ 
\hline 
$\alpha_1$ & 4 & 4.00 & 3.96 & (3.41, 4.82) & 3.82 & 3.82 & (3.32, 4.32) \\ 
%\hline 
$\alpha_2$ & 8 & 8.80 & 8.78 & (7.14, 10.53) & 8.85 & 8.84 & (7.02, 10.37) \\ 
%\hline 
$\beta_1$ & 2 & 1.99 & 1.99 & (1.82, 2.15) & 2.05 & 2.06 & (1.92, 2.18) \\ 
%\hline 
$\beta_2$ & 8 & 7.95 & 8.04 & (6.44, 8,67) & 8.09 & 8.09 & (7.60, 8.54) \\ 
%\hline 
$\omega_1$ & 2/3 & 0.67 & 0.68 & (0.52, 0.73) & 0.70 & 0.70 & (0.66, 0.74) \\ 
%\hline 
$\omega_2$ & 1/3 & 0.33 & 0.32 & (0.27, 0.48) & 0.30 & 0.30 & (0.26, 0.34) \\ 
%\hline 
$\theta$ & 9 & 9.02 & 9.02 & (8.95, 9.08) & 8.63 & 8.67 & (8.82, 9.02) \\ 
%\hline 
$\xi$ & 0.4 & 0.46 & 0.45 & (0.21, 0.78) & 0.37 & 0.35 & (0.11, 0.67) \\ 
%\hline 
$\sigma$ & 2 & 1.98 & 1.97 & (1.71, 2.26) & 1.98 & 1.98 & (1.71, 2.26) \\ 
\hline 
\end{tabular}
\caption{Statistics of the posterior distributions under the prior based on losses (KL prior) and under the uniform prior.} \label{tab:simulation}
\end{table}
We can see that the true parameter values are within the limits of the $95\%$ credible intervals of the respective posterior. It is not possible to complete a thorough comparison between the proposed priors on the basis of one sample only, and we will be performing this exercise in the next section. However, focusing on the GPD parameters, it appears that the two priors have similar performances when the credible intervals are considered. Finally, for the same model, we have considered an increased sample size of $n=5,000$, and the result was to obtain narrower credible intervals for all the parameters (not shown here). It is in fact appropriate to expect this result as the likelihood function, for a relatively large data set, has sufficient information to identify the true parameter values of the model, and it is also for this reason that a likelihood representing the whole model is appropriate.
\begin{figure}[hbtp]
     \begin{center}
        \subfigure{%
            \label{sim_1}
            \includegraphics[width=0.3\textwidth]{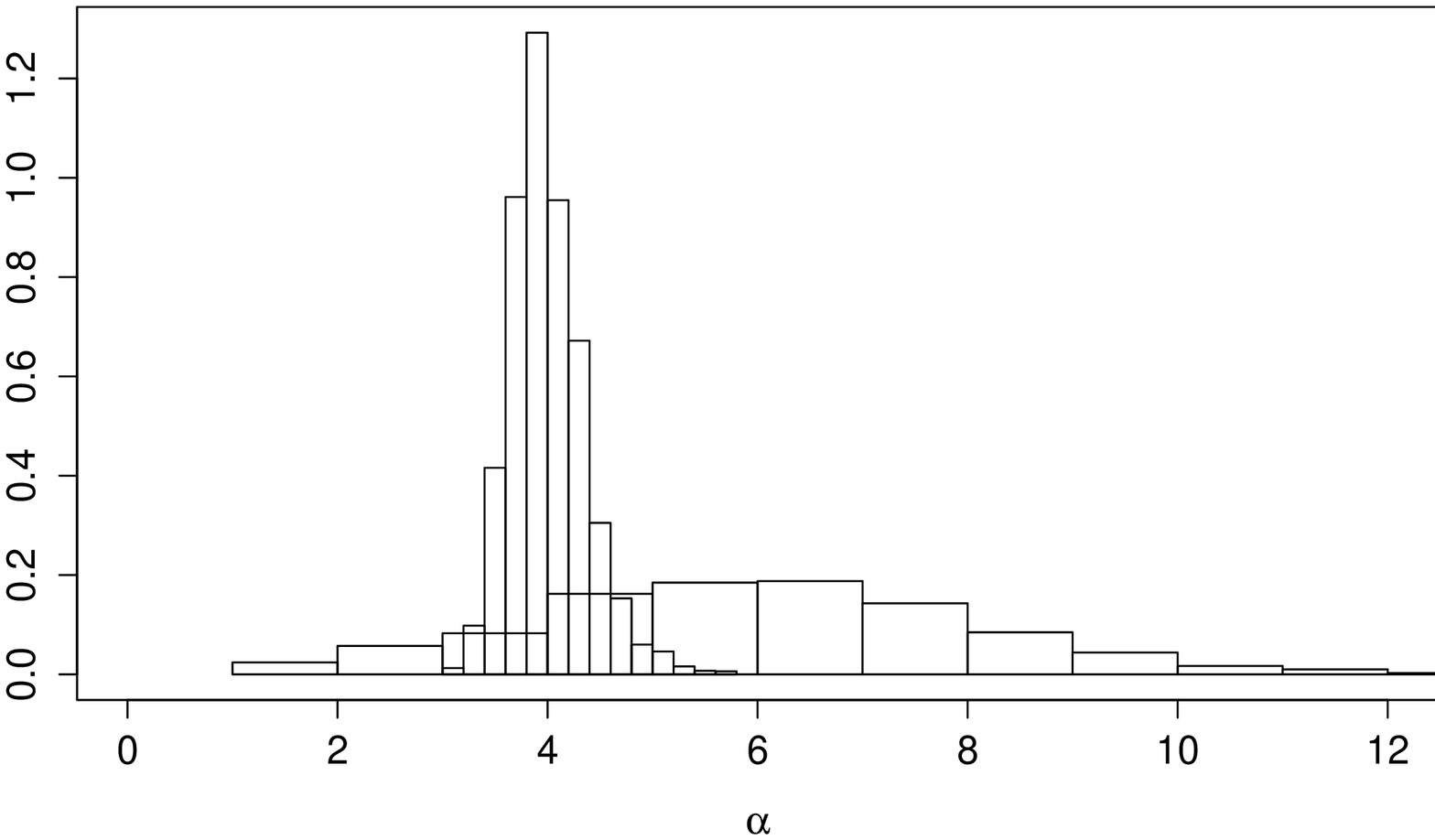}
        } %
        \subfigure{%
           \label{sim_2}
           \includegraphics[width=0.3\textwidth]{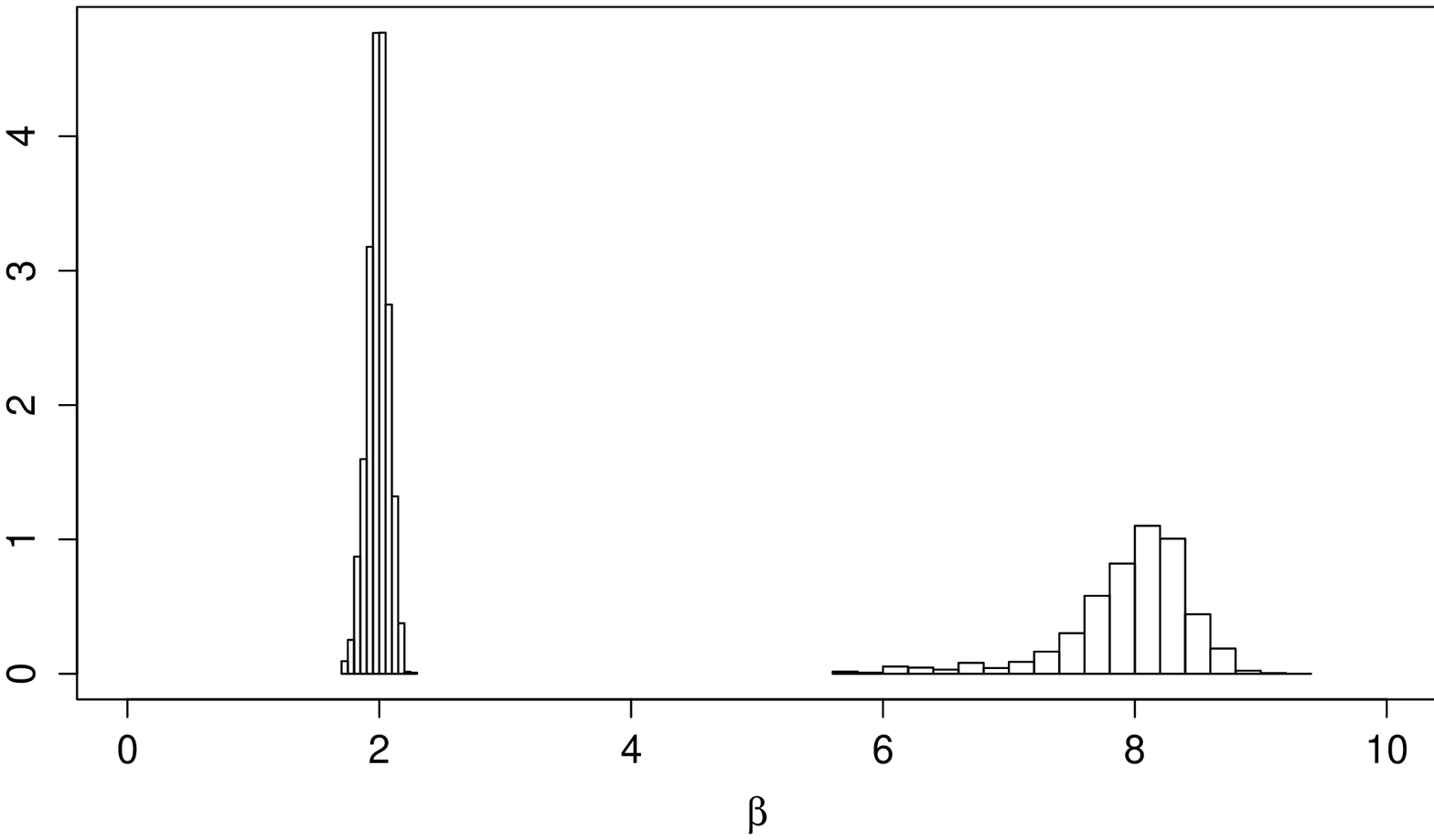}
        } %  
        \subfigure{%
            \label{sim_3}
            \includegraphics[width=0.3\textwidth]{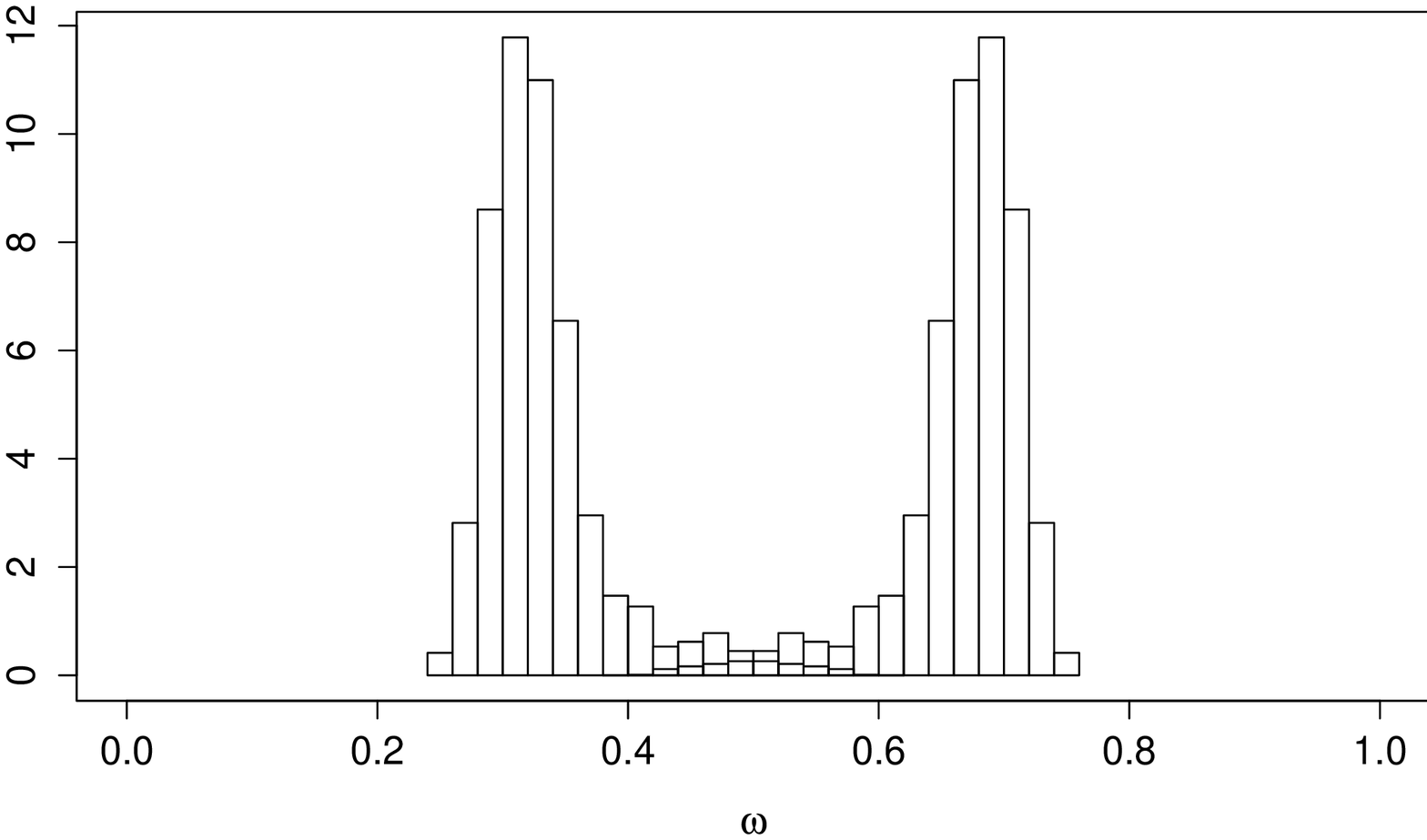}
        } %
        \subfigure{%
            \label{sim_4}
            \includegraphics[width=0.3\textwidth]{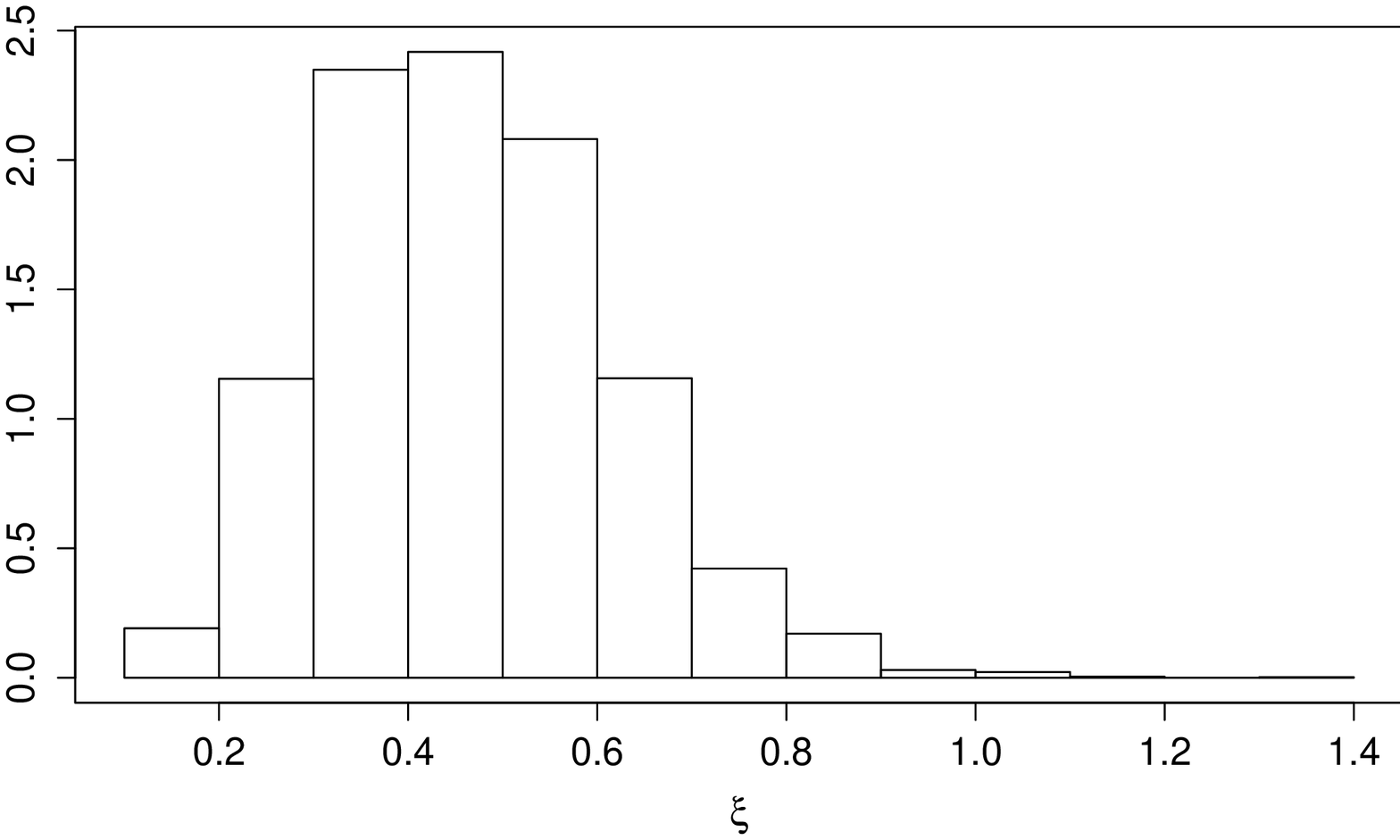}
        } %
        \subfigure{%
            \label{sim_5}
            \includegraphics[width=0.3\textwidth]{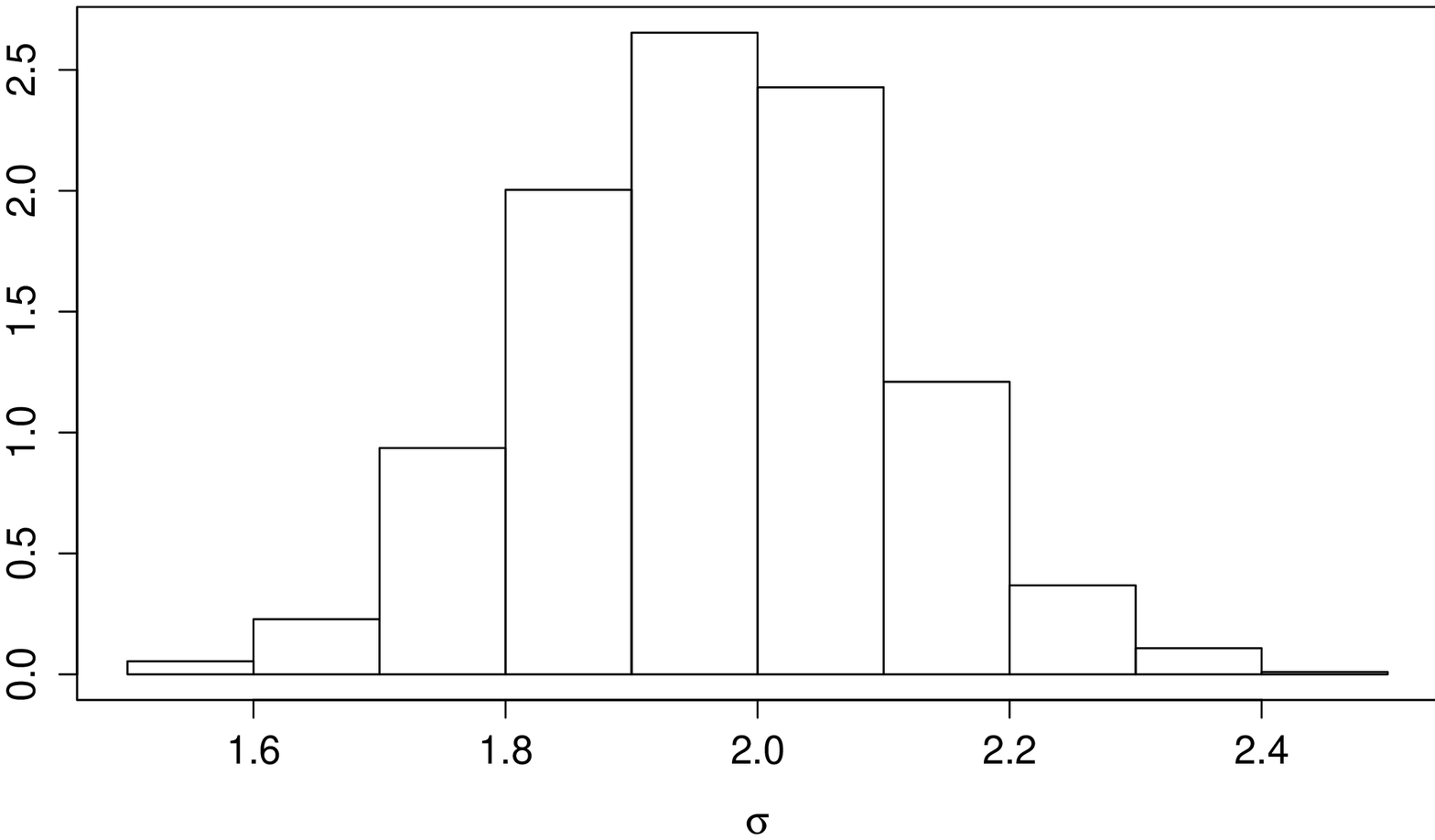}
        } %
        \subfigure{%
            \label{sim_6}
            \includegraphics[width=0.3\textwidth]{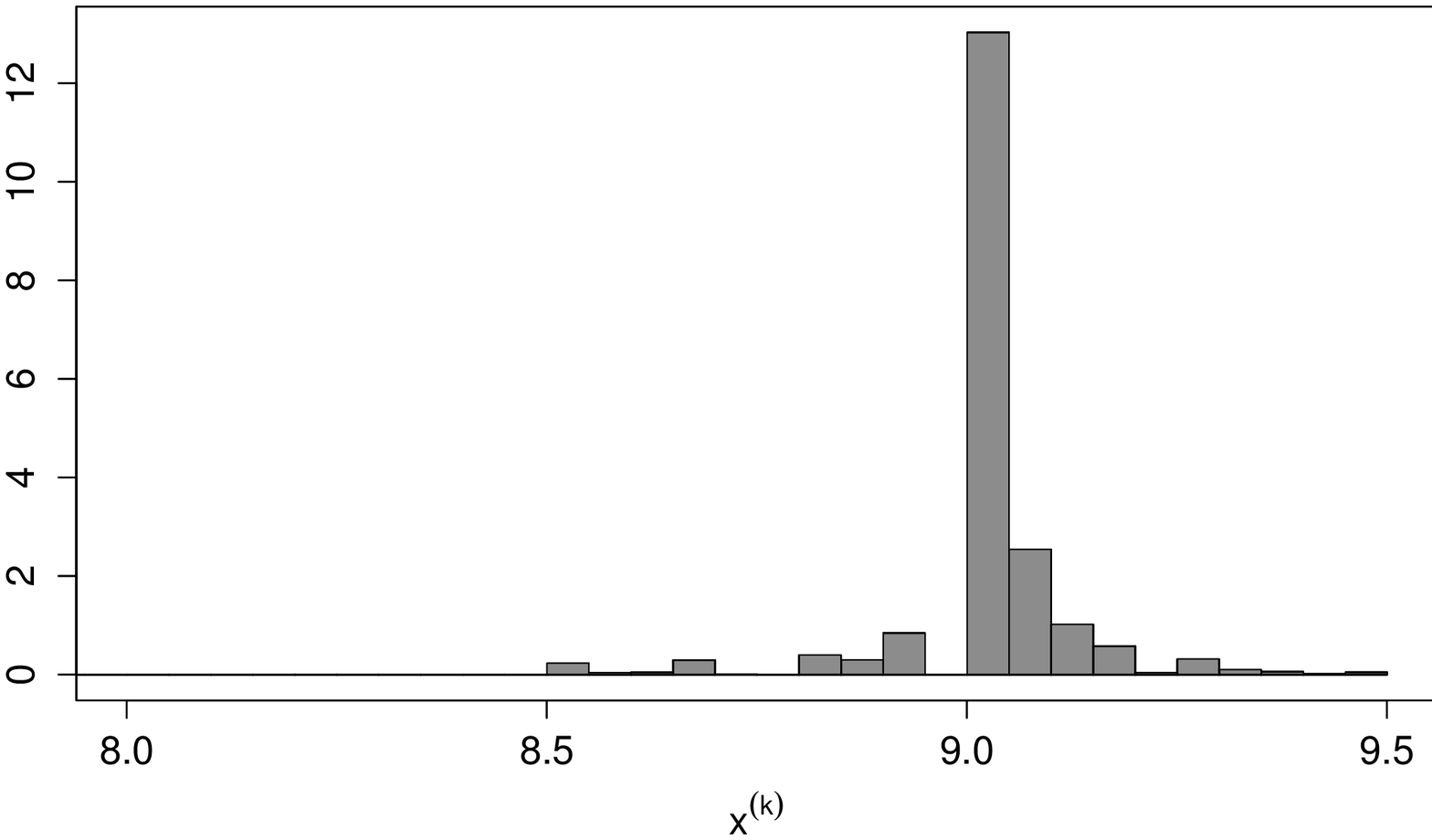}
        } %
    \end{center}
    \caption{%
        Histograms of the posterior distributions of the parameters of the simulated model, when the prior based on losses for $x^{(k)}$ is considered. In the top row, from left to right, we have the histograms of the shape parameters $\alpha_1$ and $\alpha_2$ (left plot), of the means $\beta_1$ and $\beta_2$ of the means of the gamma densities (middle plot)  and of the weights $\omega_1$ and $\omega_2$ (right plot). In the bottom row we have the histograms of the posterior distributions of the parameters of the GPD: $\xi$ (left plot), $\sigma$ (middle plot) and $x^{(k)}$ (right plot).
     }%
   \label{fig:simulhist}
\end{figure}
Figure \ref{fig:simulhist} shows the histograms of the posterior distributions of the parameters when the prior based on losses is employed; we have omitted the analogous graphs when the uniform prior is considered as they did not show any worthwhile difference. In the top row we have the parameters of the mixture representing the bulk data. To increase readability we have grouped in a single plot the histograms of the same parameter of each mixture component. That is, top row from left to right, we have the shapes parameters $\beta$'s, the means $\alpha$'s and the weights $\omega$ of the components. The bottom row is dedicated to the parameters of the GPD. The histogram of most interest is the one on the posterior of the threshold $x^{(k)}$. We note that, due to the discrete nature of the distribution, i.e. on the order statistics, it lacks of smoothness. This is expected as the observations will not cover the whole space of $x^{(k)}$ and some order statistics may correspond to contiguous values with different spacing.

\subsection{Frequentist performances of the yielded posterior distributions}
The aim of the simulation study presented in this section is to analyse the performances of the proposed discrete priors on the order statistics by obtaining two frequentist statistics on repeated samples across a variety of model scenarios: the coverage of the $95\%$ credible interval of the posterior distribution and the mean squared error (MSE) from the mean, of the posterior distribution. As the focus of our work is mainly on the threshold of the GPD, we have kept the model structure fixed, in the sense that the mixture for the bulk data has two components (gamma densities) for all the sampling cases. The changes were in the parameters of the GPD and the sample size. In detail, we have considered two sample sizes, that is $n=1,000$ and $n=5,000$. For the shape parameter of the GPD, we have set $\xi=\{0.4,0.8,1.0,2.0,3.0,4.0\}$, whilst for the scale parameter we have chosen $\sigma=\{2,4\}$. In order to avoid a tedious illustration of the results, we show the simulations for the case with $\sigma=2$ only, as we have not identified any notable difference in the simulations with $\sigma=4$. Finally we have set the threshold at $\theta=7$ and at $\theta=9$.

\begin{figure}[hbtp]
     \begin{center}
        \subfigure{%
           \label{Cov_u7_n1000}
           \includegraphics[scale=0.33]{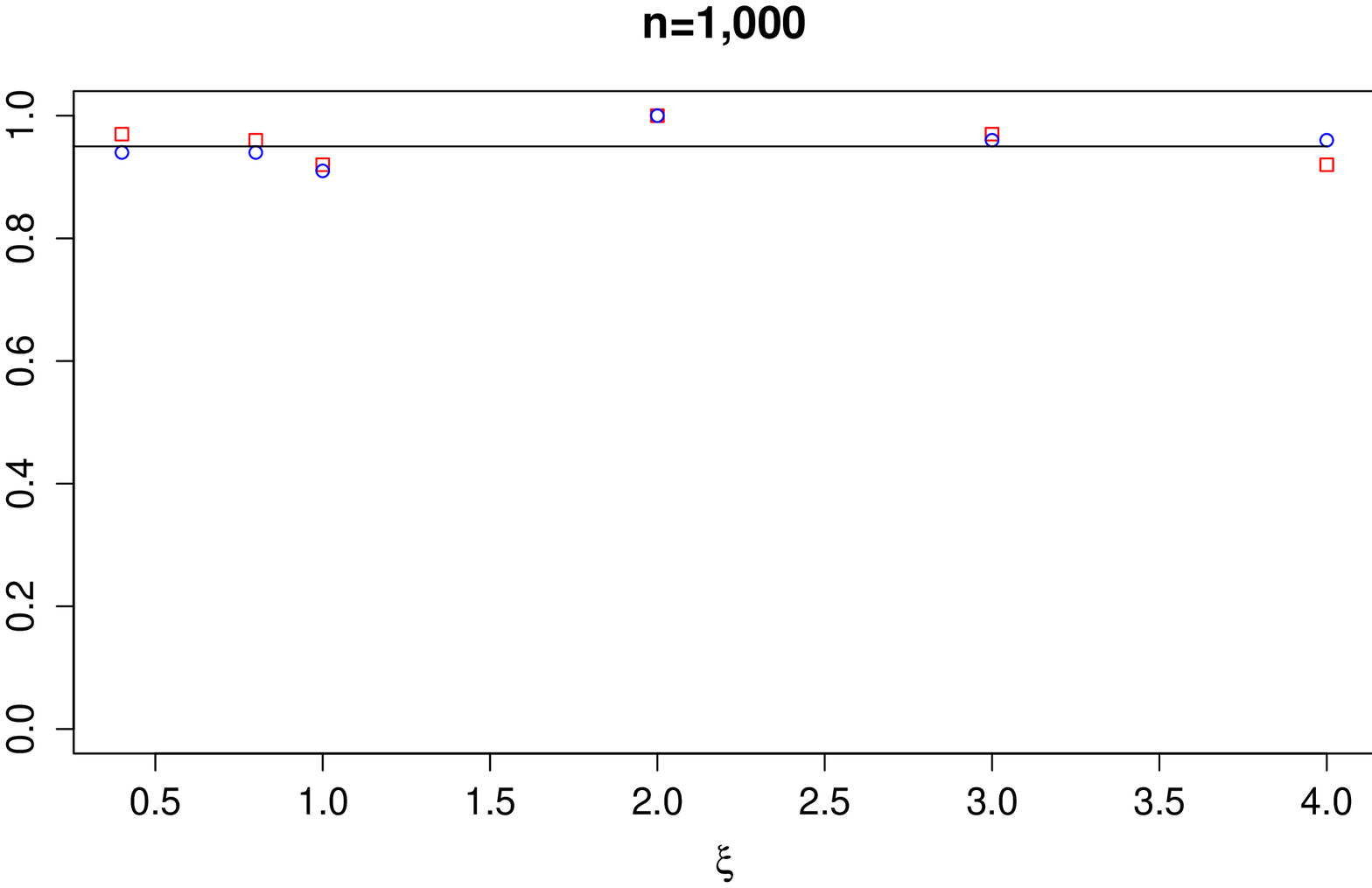} 
        } %
        \subfigure{%
           \label{Cov_u7_n5000}
           \includegraphics[scale=0.33]{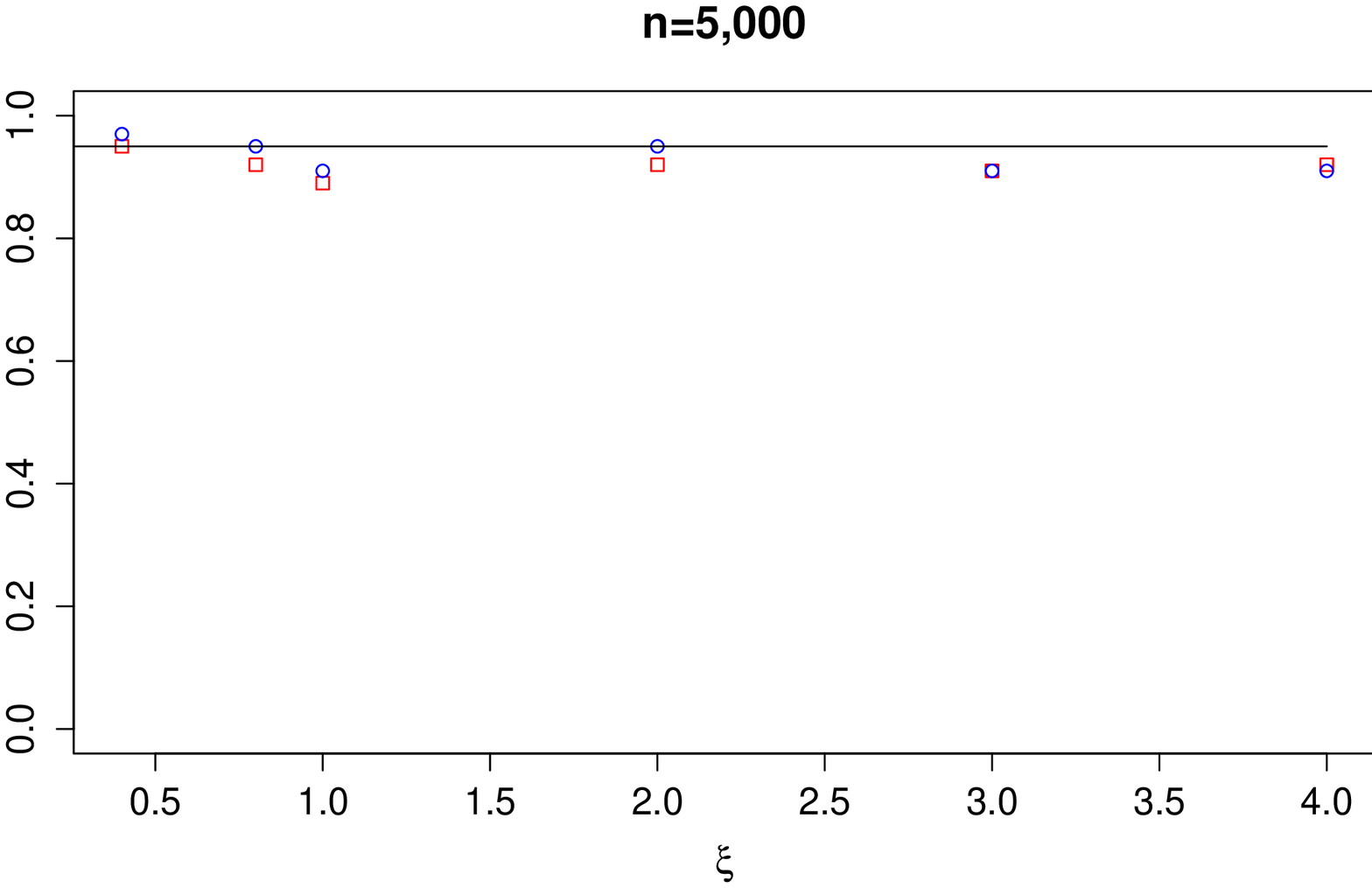} 
        } %  
        \subfigure{%
           \label{Mse_u7_n1000}
           \includegraphics[scale=0.50]{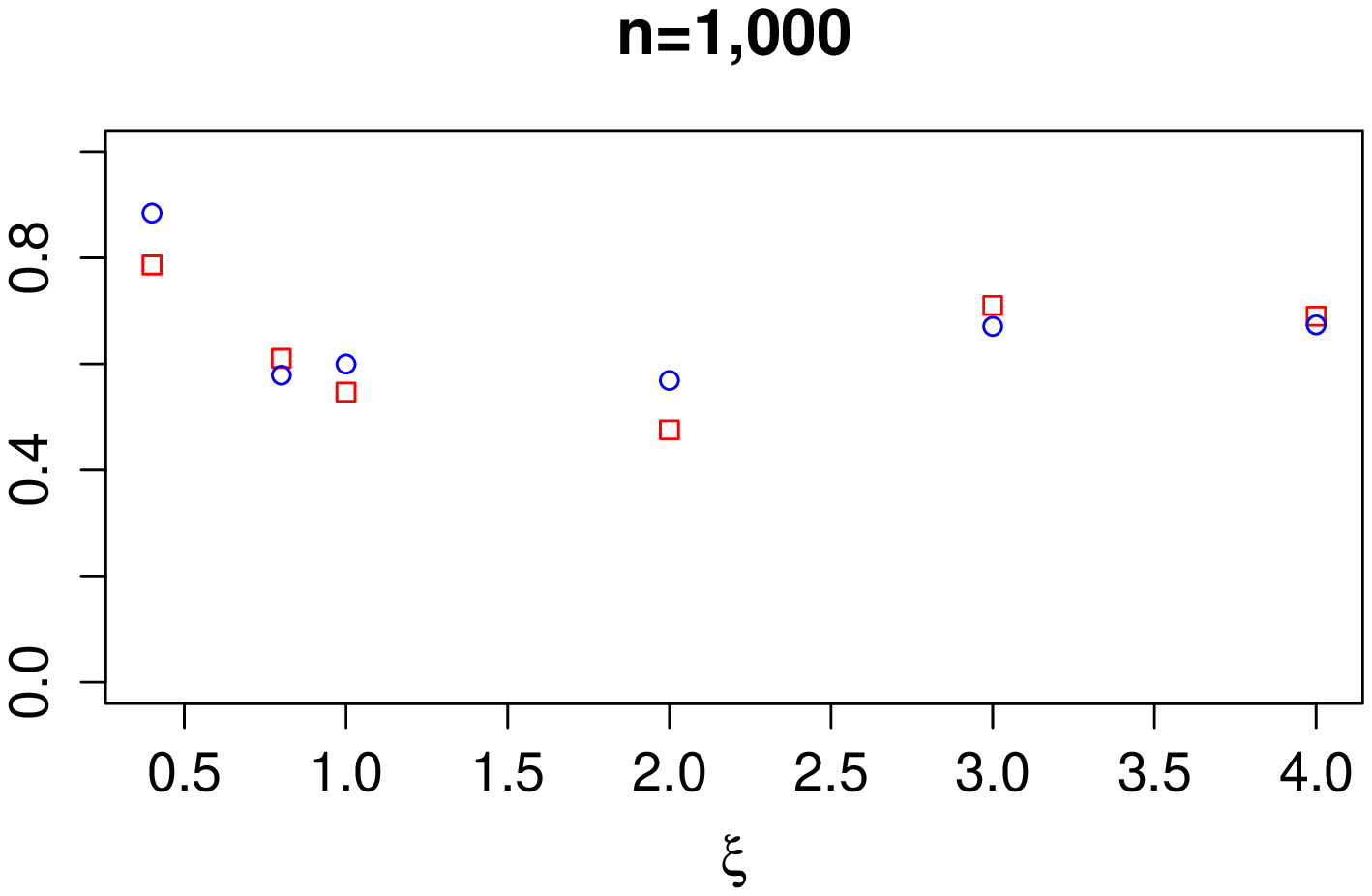} 
        } %
        \subfigure{%
           \label{Mse_u7_n5000}
           \includegraphics[scale=0.33]{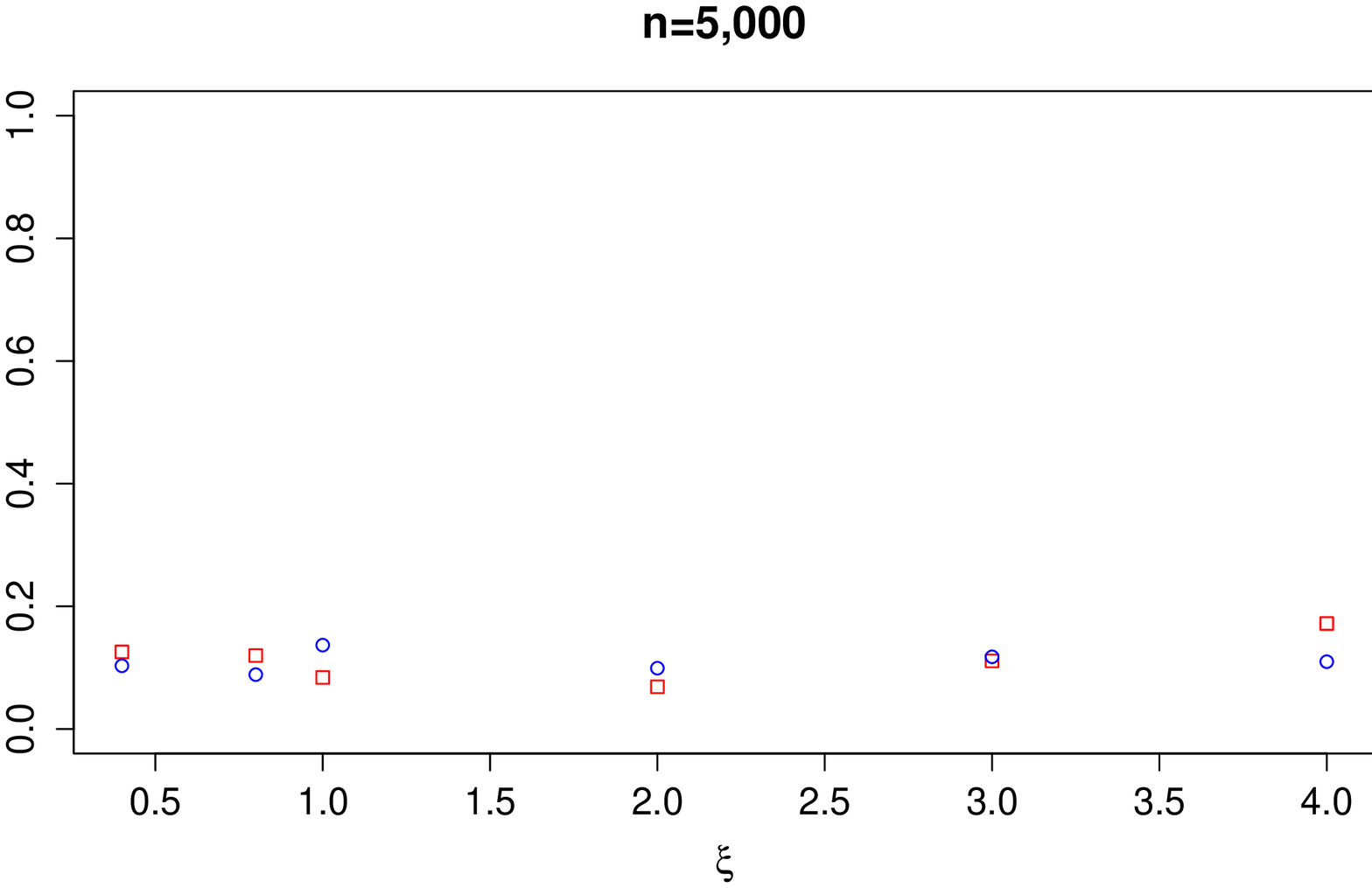} 
        } %
    \end{center}
    \caption{%
        Coverage of the $95\%$ credible interval of the posterior for $\theta=7$ for $n=1,000$ (top left) and for $n=5,000$ (top right), and MSE from the mean for $n=1,000$ (bottom left) and for $n=5,000$ (bottom right). Each graph shows the results from the uniform prior (blue circle) and the prior based on losses (red square).
     }%
   \label{fig:simu_freq_u7}
\end{figure}
\begin{figure}[hbtp]
     \begin{center}
        \subfigure{%
           \label{Cov_u9_n1000}
           \includegraphics[scale=0.33]{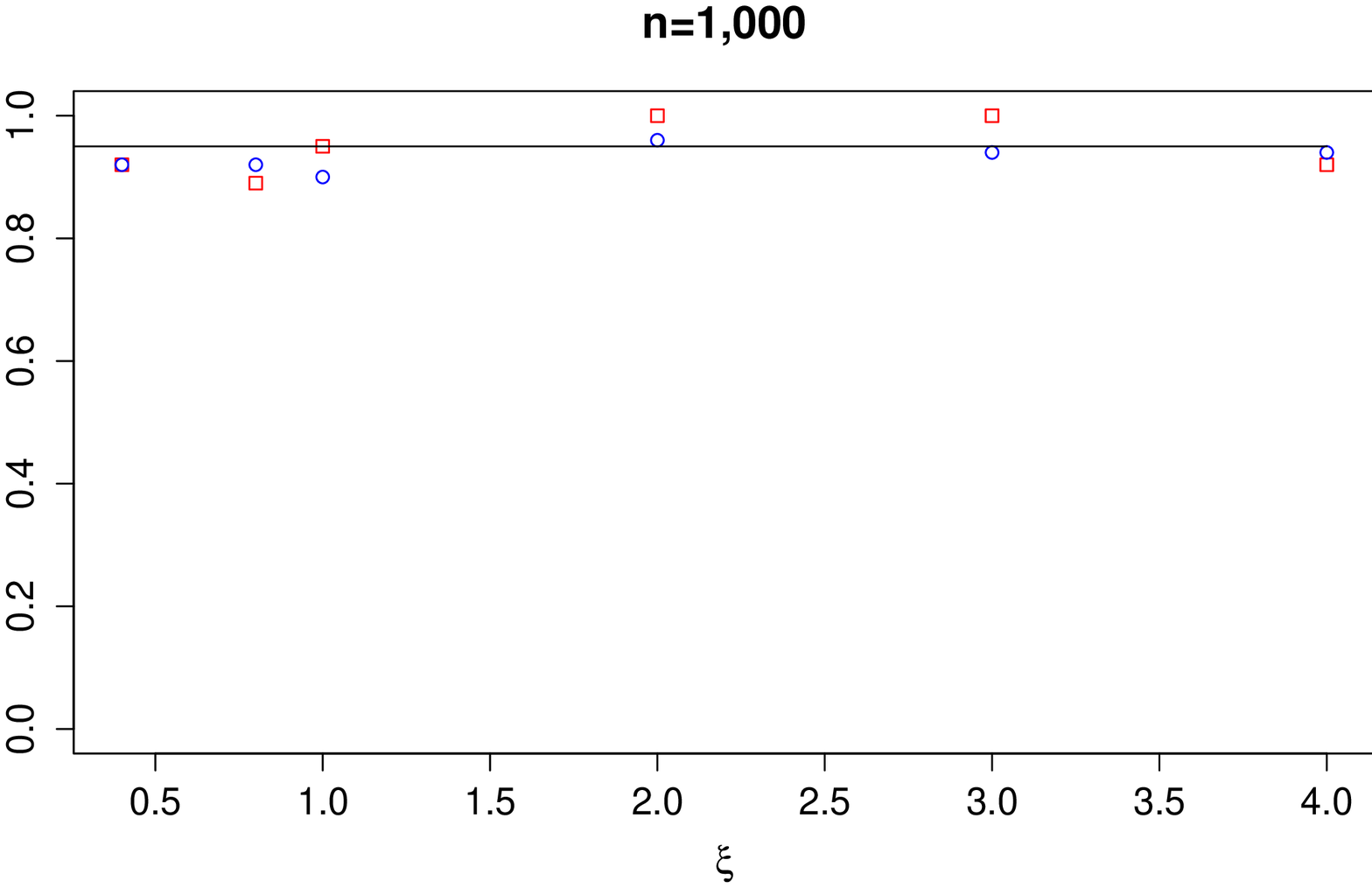} 
        } %
        \subfigure{%
           \label{Cov_u9_n5000}
           \includegraphics[scale=0.33]{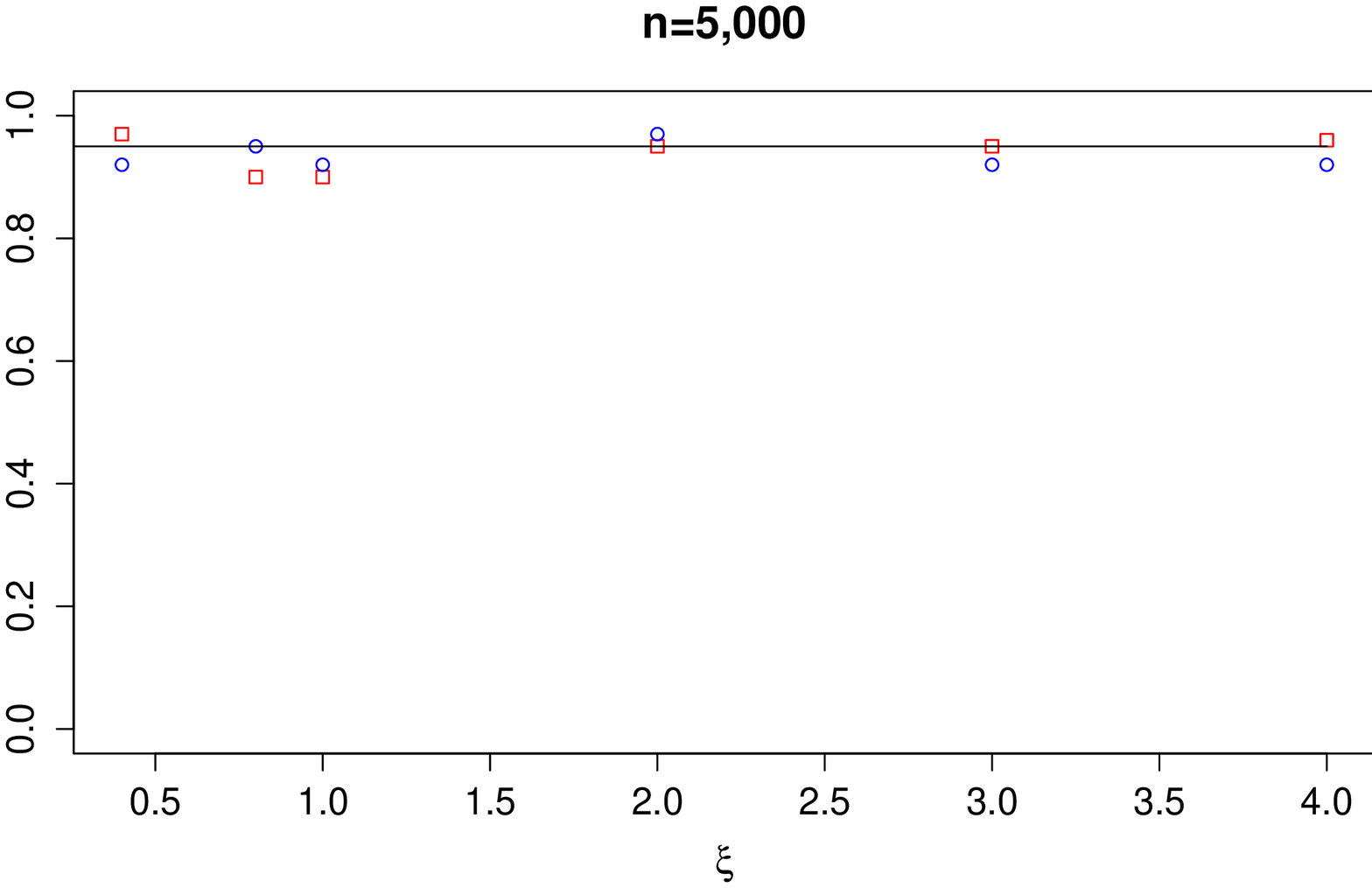} 
        } %  
        \subfigure{%
           \label{Mse_u9_n1000}
           \includegraphics[scale=0.33]{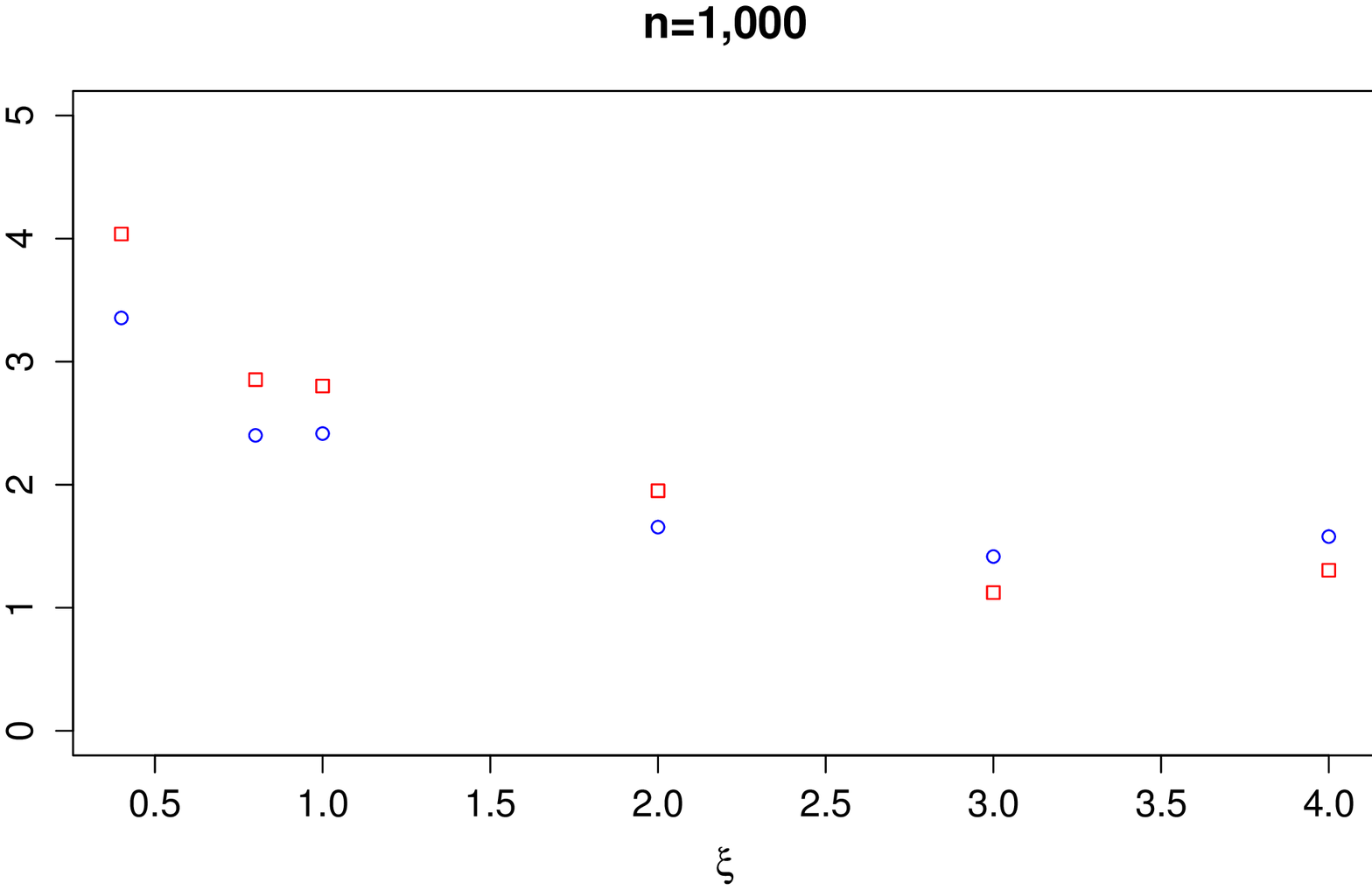} 
        } %
        \subfigure{%
           \label{Mse_u9_n5000}
           \includegraphics[scale=0.33]{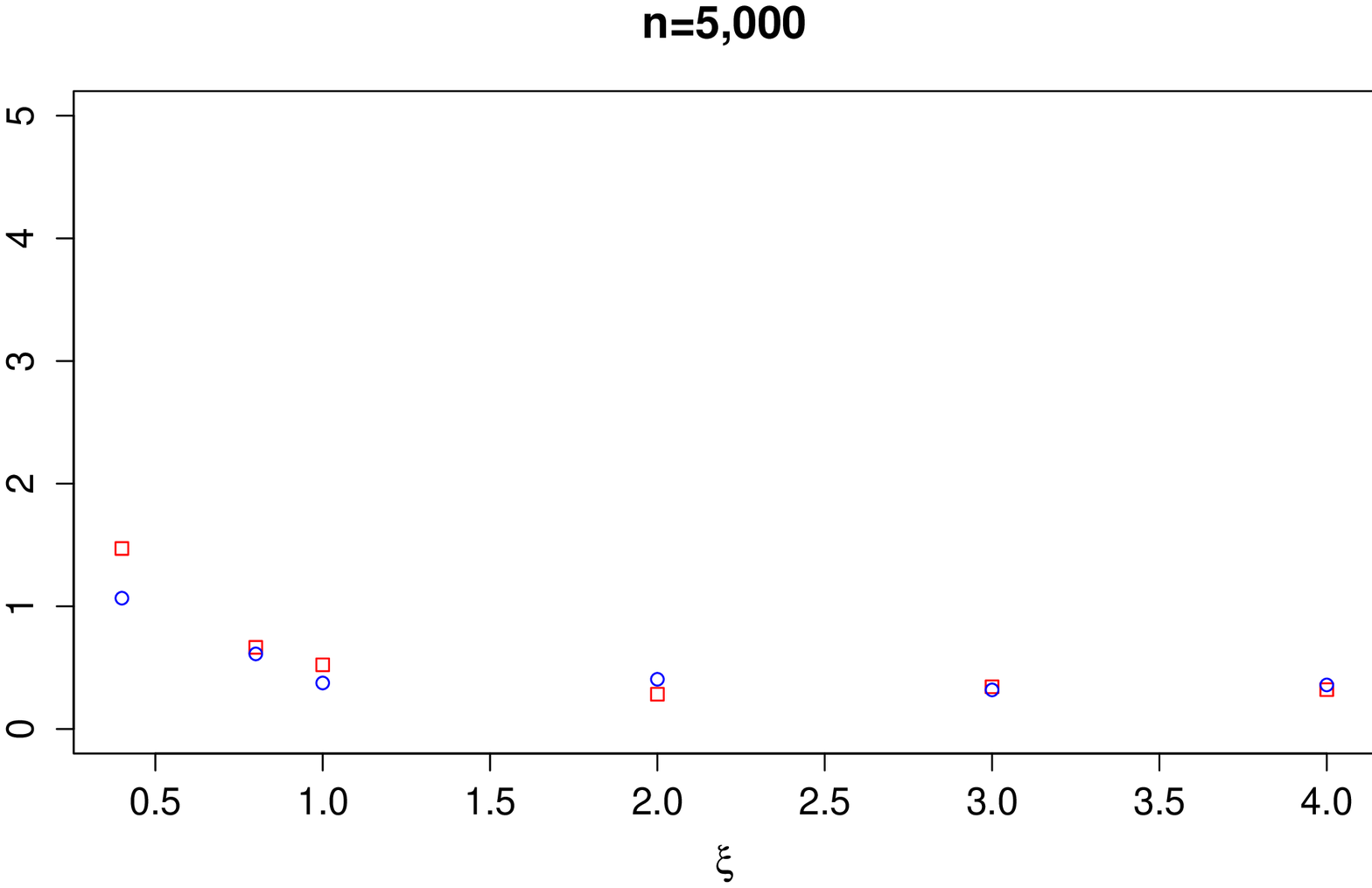} 
        } %
    \end{center}
    \caption{%
        Coverage of the posterior for $\theta=9$ for $n=1,000$ (top left) and for $n=5,000$ (top right), and MSE from the mean for $n=1,000$ (bottom left) and for $n=5,000$ (bottom right). Each graph shows the results from the uniform prior (blue circle) and the prior based on losses (red square).
     }%
   \label{fig:simu_freq_u9}
\end{figure}

Figures \ref{fig:simu_freq_u7} and \ref{fig:simu_freq_u9} illustrate the comparison of the frequentist performance of the uniform prior and the prior based on losses as in \eqref{eq_prior_5}. The coverage of the $95\%$ credible interval of the posterior for $x^{(k)}$, for both $\theta=7$ and $\theta=9$, is compatible with the nominal value and appears to be unaffected by the value of the shape parameter $\xi$ and by the sample size $n$. To analyse the MSE from the mean, let us first consider the case $\theta=7$. As one would expect, the MSE is smaller for larger sample sizes. It appears that there is larger variability in its estimate for different values of $\xi$ when $n=1,000$ then when $n=5,000$. A similar behaviour can be seen in the case the threshold is set equal to 9.
In addition, when we compare the MSE for $\theta=7$ and $\theta=9$, we note that its value is higher in the second case. Given that the rest of the mixture model is kept unchanged, a higher threshold implies less data included in the GPD part of the likelihood; therefore, less information to estimate the parameters. When comparing the two discrete priors for the threshold, it seems that the overall frequentist properties are reasonably similar, especially for larger sample sizes. For the smaller sample size $n=1,000$, we note different performances in the lower end on the parameter space of $\xi$ when the threshold is set to 7. With a threshold of the GPD equal to 9, it appears that the uniform priors outperforms the prior based on losses for low values of $\xi$, but it is outperformed for growing values of the shape parameter. In any case, the differences observed appear to be restrained.

%--- REAL DATA ------------------------------------------------
\section{Real data modeling}\label{sc_real}
In this section we show the application of the proposed discrete prior distributions for the threshold of the GPD. The first example is an application from insurance and we analyse the popular data set of losses due to fires in Denmark over a decade. In the second example we analyse financial data (NASDAQ-100 returns) and we show that the proposed priors, and the overall model, allow for the information about the threshold that is contained in the bulk data to be taken into account.

\subsection{An application from insurance}
In the first application of the proposed discrete priors for the GPD threshold we analyse the popular Danish fire loss data. This data set has been largely analysed in the literature, including \cite{Mcneil:1997}, \cite{Frigetal:2002} and \cite{Castel:2011}, and it reports 2167 insurance losses deriving from as many industrial fires occurred in Denmark over the period 1980 to 1990. The losses are valued in millions of Danish krone (DKK) adjusted to the year 1985 values, for comparison purposes.
\begin{figure}[hbtp]
	\begin{center}
		\includegraphics[scale=0.5]{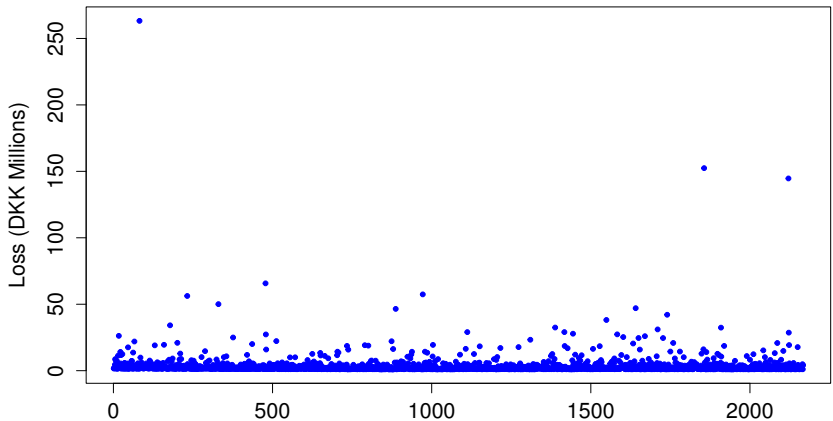}
	\end{center}
	\caption{Danish fire loss observations in chronological order.}
	\label{fig:danish_hist}
\end{figure}
Figure \ref{fig:danish_hist} shows the data in chronological order, say $y$, where it is possible to see that the majority of observations are grouped below the value of DKK 25 millions, with an increasing sparsity the more the loss amount becomes extreme. As such, it appears to be appropriate to model the quantity by a mixture model with a bulk data component and a GPD to represent the heavy-tailed behaviour.

\begin{figure}[hbtp]
     \begin{center}
        \subfigure{%
           \label{Danish_prior_1}
           \includegraphics[scale=0.33]{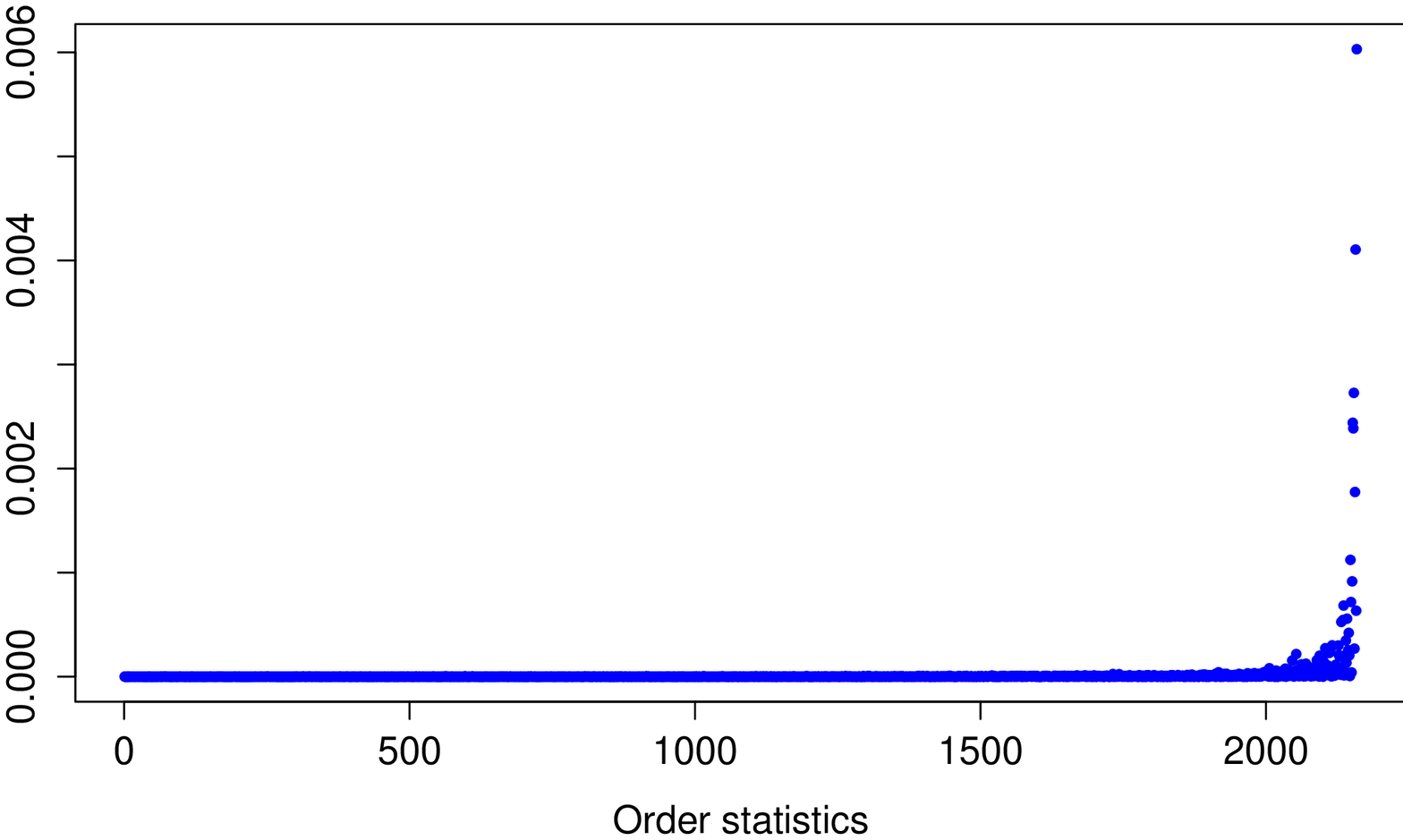} 
        } %
        \subfigure{%
           \label{Danish_prior_2}
           \includegraphics[scale=0.33]{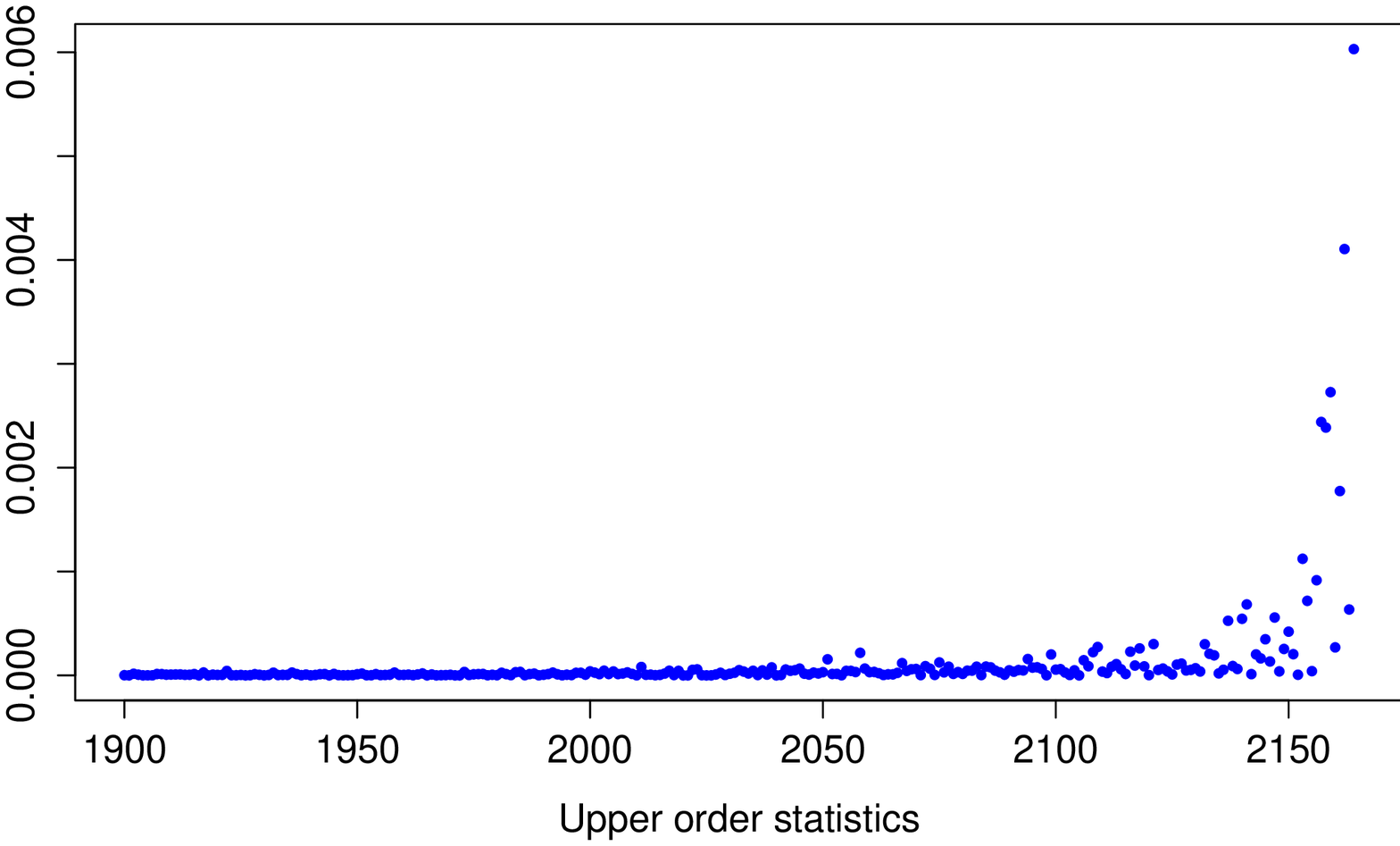} 
        } %  
    \end{center}
    \caption{%
        Discrete prior distribution for the the threshold $x^{(k)}$ based on losses. The left graph represents the normalised prior for the whole data set, the right graph the normalised prior for the upper order statistics only - i.e. from $y^{(1900)}$. The prior has been drawn by setting $\xi=0.583$ and $\sigma=5.921$.
     }%
   \label{fig:danish_prior}
\end{figure}
Whilst the uniform prior for the threshold is easy to picture, it is appropriate to have a feeling of what the proposed prior based on losses may look like. As discussed in Section \ref{sc_thetaprior}, the prior based on losses depends on the values of $\xi$ and $\sigma$. Thus, for illustration purposes only, we set the two parameter at the estimated values in \cite{Castel:2011}; in particular, to the posterior medians $\xi=0.583$ and $\sigma=5.921$. In Figure \ref{fig:danish_prior} we illustrate the prior given the above two values for $\xi$ and $\sigma$, where on the left graph we have the prior for the order statistics corresponding to all the data, and on the right graph the prior for the upper order statistics only, i.e. from $y^{(1900)}$ to allow for a better visualisation of the right tail of the prior distribution. We note that the behaviour of the prior is consistent with what expected and with the results in the simulation exercises; that is, the prior probability on the lower order statistics is almost uniform and it grows for the upper order statistics.

The first part of the analysis was to identify the most appropriate model to represent the data; in particular, the number of component in the mixture for the bulk data. This has been accomplished by running the Monte Carlo simulation for $r=1,2,3,4,5,6$, under each proposed discrete prior for $x^{(k)}$, and selecting the model for which none of the weights $\omega$ converged to zero. Under both priors for the threshold, the selected model was for $r=3$. Therefore, the model chosen to represent the Danish fire loss data is of the form
\begin{equation*}\label{eq_danishmodel}
f(y|\gamma,\xi,\sigma,x^{(k)}) = 
	\begin{cases}
	\sum_{j=1}^3 \omega_j f_j(y|\alpha_j,\beta_j) & y<x^{(k)} \\
	[1-H(x^{(k)}|\gamma)]g(y|\xi,\sigma,x^{(k)}) & y\geq x^{(k)},
					\end{cases}
\end{equation*}
where $\gamma=\{\omega_1,\omega_2,\omega_3,\alpha_1,\alpha_2,\alpha_2,\beta_1,\beta_2,\beta_3\}$, and $H(x^{(k)}|\gamma)$ is the value of the cumulative distribution function of the bulk part of the model evaluated at $x^{(k)}$.

\begin{table}
\centering
\begin{tabular}{|c|ccc|ccc|}
\hline 
 & \multicolumn{3}{c|}{KL Prior} & \multicolumn{3}{c|}{Uniform Prior} \\ 
\hline 
Parameters & Mean & Median & $95\%$ C.I. & Mean & Median & $95\%$ C.I. \\ 
\hline 
$\alpha_1$ & 33.83 & 34.06 & (31.78,36.04) & 32.83 & 32.62 & (29.04,36.20) \\ 
%\hline 
$\alpha_2$ & 14.62 & 14.42 & (12.43,17.94) & 15.89 & 15.83 & (13.06,19.00) \\ 
%\hline 
$\alpha_3$ & 4.92 & 4.85 & (3.82,6.56) & 6.81 & 6.56 & (4.70,7.73) \\ 
%\hline 
$\beta_1$ & 1.31 & 1.31 & (1.28,1.34) & 1.31 & 1.31 & (1.27,1.34) \\ 
%\hline 
$\beta_2$ & 2.03 & 2.02 & (1.92,2.15) & 2.00 & 1.97 & (1.84,2.11) \\ 
%\hline 
$\beta_3$ & 5.00 & 5.00 & (4.62,5.40) & 4.54 & 4.53 & (4.16,5.46) \\ 
%\hline 
$\omega_1$ & 0.38 & 0.38 & (0.33,0.43) & 0.38 & 0.38 & (0.31,0.43) \\ 
%\hline 
$\omega_2$ & 0.34 & 0.34 & (0.29,0.40) & 0.33 & 0.33 & (0.28,0.39) \\ 
%\hline 
$\omega_3$ & 0.28 & 0.28 & (0.24,0.32) & 0.29 & 0.29 & (0.25,0.34) \\ 
%\hline 
$\theta$ & 5.79 & 5.79 & (4.93,7.54) & 5.16 & 4.45 & (4.08,7.99) \\ 
%\hline 
$\xi$ & 0.53 & 0.52 & (0.32,0.78) & 0.64 & 0.65 & (0.37,0.91) \\ 
%\hline 
$\sigma$ & 5.20 & 5.18 & (4.04,6.60) & 4.02 & 3.23 & (2.32,6.54) \\ 
\hline 
\end{tabular} 
\caption{Summary statistics of the posterior distributions for the mixture model for the Danish fire loss dataset.}\label{tab: danish}
\end{table}

The inferential results are detailed in Table \ref{tab: danish}. We compare the estimates of the GPD parameters with the values obtained by \cite{Castel:2011}, which are 5.30 for the threshold, 0.58 for the shape parameter and 5.92 for the scale parameter. The results obtained by using both the discrete priors on the order statistics appears to be in accordance with the values estimated by \cite{Castel:2011}. There is an exception in the estimate of $\sigma$ when the uniform prior on the order statistics is employed; however, the estimated value is well within the credible interval. If we compare the estimates obtained by using the two proposed priors, we note that there is strong agreement in the results for what it concerns the bulk part of the model. Besides a slightly larger credible interval for the estimate of the scale parameters under the uniform prior, it appears that there are no estimates notably different to be highlighted.

\subsection{An application from finance}
In the second example we analyse the daily increments for the NASDAQ-100 stock index. In particular, we consider the adjusted closing price from the 2$^{\textnormal{nd}}$ of January 1985 to the 31$^{\textnormal{st}}$ of May 2002, for a total of $n = 4394$ observations. This data set has been analysed by \cite{Behr:2004} and we are able to compare our results with theirs.

\begin{figure}[hbtp]
	\begin{center}
		\includegraphics[scale=0.5]{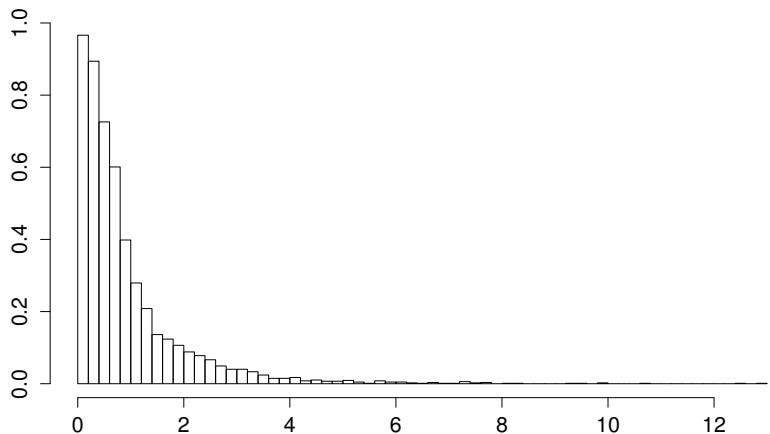}
	\end{center}
	\caption{Histogram of the NASDAQ-100 daily increments}
	\label{fig:nasdaq_hist}
\end{figure}

\begin{figure}[hbtp]
     \begin{center}
        \subfigure{%
           \label{Nasdaq_prior_1}
           \includegraphics[scale=0.33]{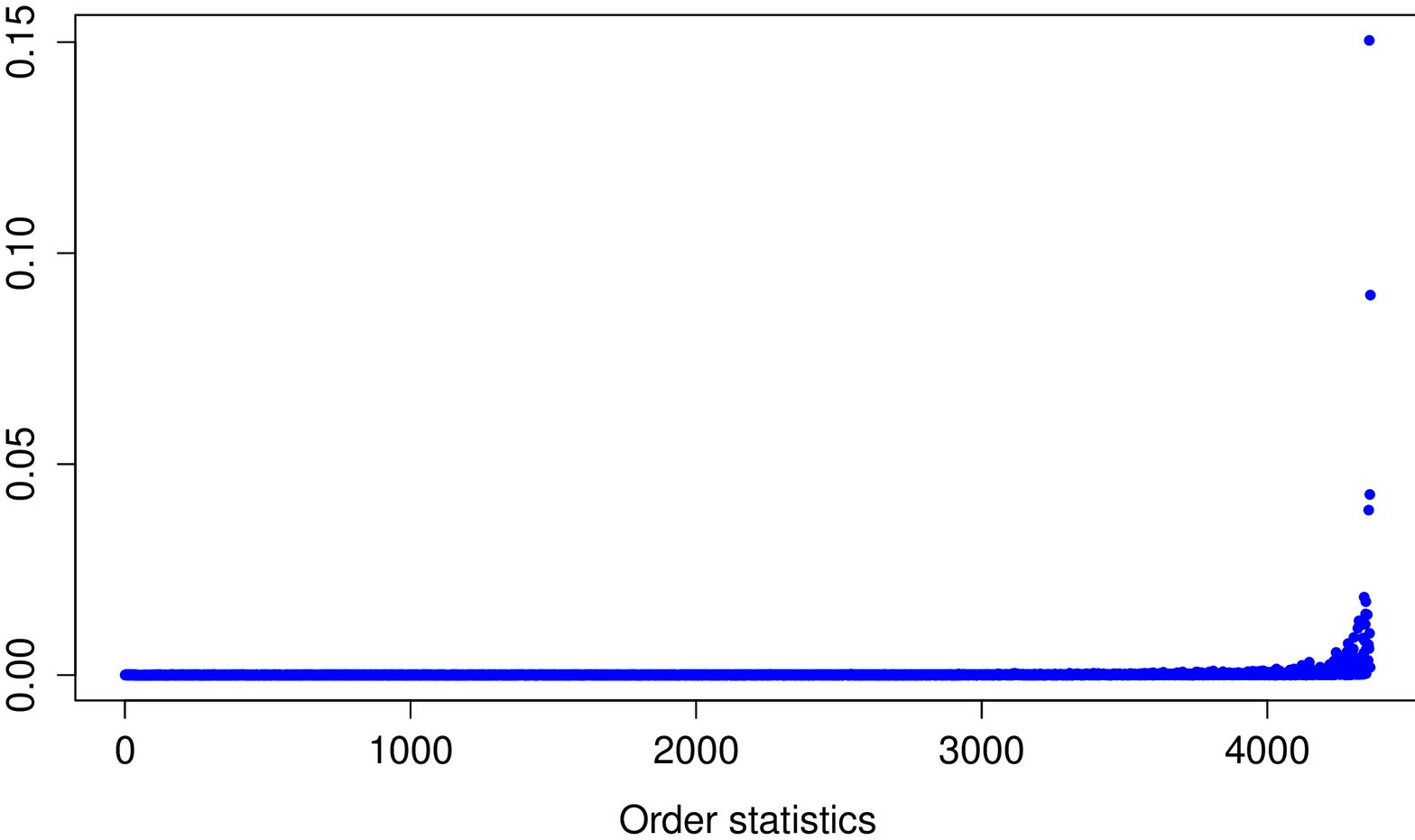} 
        } %
        \subfigure{%
           \label{Nasdaq_prior_2}
           \includegraphics[scale=0.33]{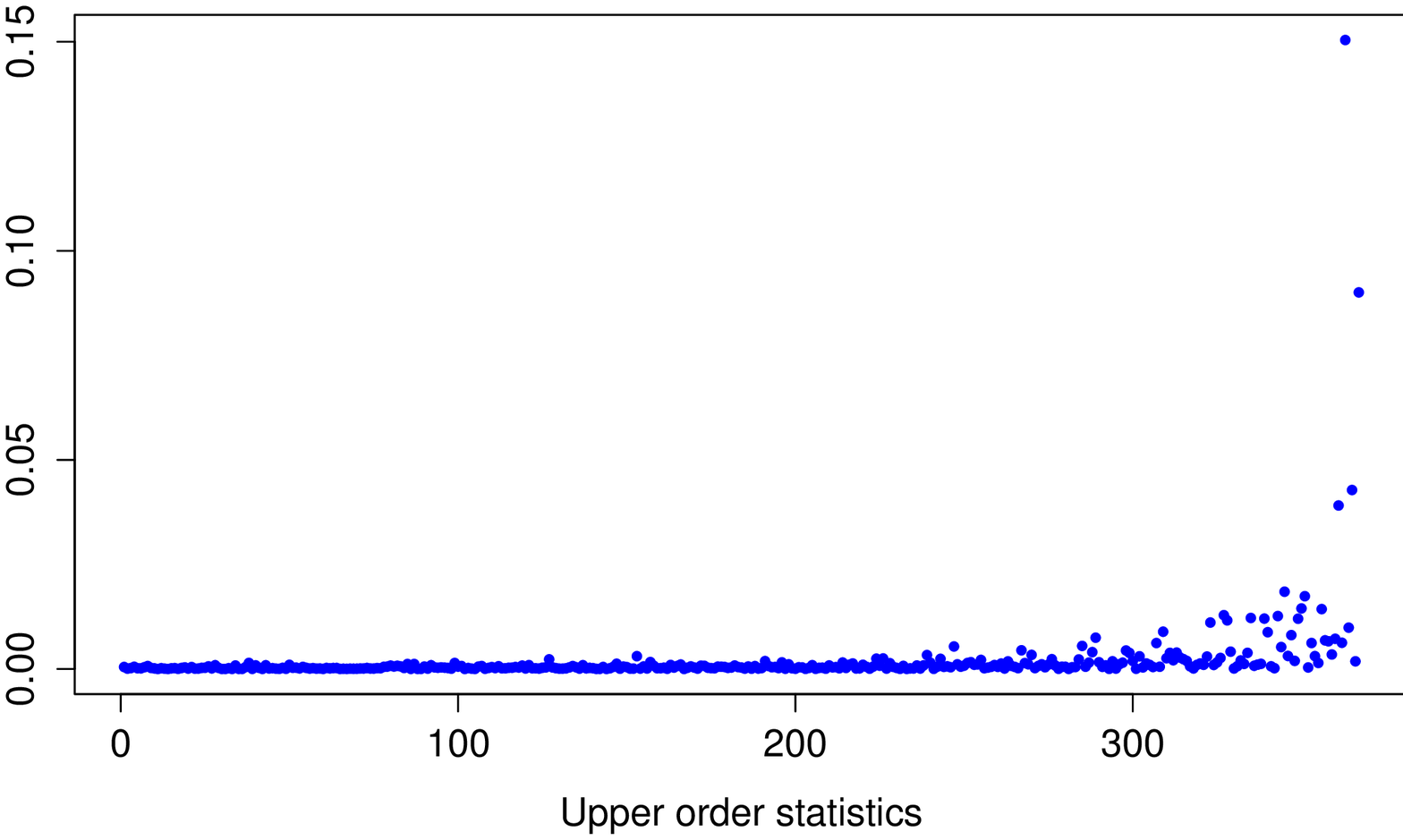} 
        } %  
    \end{center}
    \caption{%
        Discrete prior distribution for the the threshold $x^{(k)}$ based on losses, given $\xi=0.157$ and $\sigma=0.974$. The left graph represents the normalised prior for the whole data set, the right graph the normalised prior for the upper order statistics only - i.e. from $z^{(4000)}$.
     }%
   \label{fig:nasda_prior}
\end{figure}

The daily increments are obtained by applying
$$z_t = \left|\frac{r_t}{r_{t-1}}-1\right|\cdot100 \qquad t=2,\ldots,4394,$$
where $r_t$ is the adjusted closing price for the index at day $t$. The histogram of the daily increments (Figure \ref{fig:nasdaq_hist}) shows a heavy-tailed behaviour of the data, suggesting to use of a GPD to model the extreme observations.
To have a feeling of the prior distribution based on losses defined on the order statistics, we proceed as we have done for the previous example. We set the parameters to to the values estimated by the authors, that is \cite{Behr:2004}: $\xi=0.157$ and $\sigma=0.974$. Figure \ref{fig:nasda_prior} shows the prior on the whole order statistics (left graph) and the prior on the upper order statistics only (right graph).

Table \ref{tab:nasdaq_estimates} details the estimates of the GPD component. The model for the whole set of observations is always a mixture model, and we have estimated that the number of gamma densities of the mixture for the bulk data is $r=2$; however, we have considered the case where the bulk data is modelled by a mixture of gamma densities, and the case where only a single gamma distributions models the bulk data. In both cases we have estimated the GPD parameters by considering both the discrete proposed prior distributions. All the results are compared with the estimates in \cite{Behr:2004} (last column to the right).
From Table \ref{tab:nasdaq_estimates} we see that when we consider the mixture model with a single gamma distribution for the bulk data, the results we obtain by applying the uniform prior and the prior based on losses for the threshold are consistent with the estimates in \cite{Behr:2004}. 
\begin{table}
\centering
\tabcolsep=0.11cm
\begin{tabular}{|c|cc|cc|c|}
\hline 
 & \multicolumn{2}{c|}{Gamma Mixture} & \multicolumn{2}{c|}{Single Gamma} & Behrens et al. \\ 
\hline 
 & KL Prior & Uniform Prior & KL Prior & Uniform prior & Uniform prior \\ 
\hline 
$x^{(k)}$ & 1.13 (0.93, 1.18) & 1.08 (0.88, 1.11) & 0.93 (0.91, 0.94) & 0.93 (0.89, 0.96) & 0.96 (0.79, 1.13) \\ 
%\hline 
$\xi$ & 0.12 (0.01, 0.22) & 0.13 (0.01, 0.26) & 0.15 (0.08, 0.21) & 0.15 (0.09, 0.21) & 0.16 (0.09 0.23) \\ 
%\hline 
$\sigma$ & 1.07 (0.91, 3.12) & 1.01 (0.84, 3.05) & 0.98 (0.90, 1.06) & 0.98 (0.90, 1.06) & 0.97 (0.86, 1.08) \\ 
\hline 
\end{tabular}
\caption{Statistics of the posterior distributions (the $95\%$ credible intervals are in brackets) for the parameters of the GPD for the NASDAQ-100 data analysis.} \label{tab:nasdaq_estimates}
\end{table}
When we compare the estimates of the threshold of the gamma mixture for the bulk data with the ones of the single gamma mixture we note some differences. They appear to be reasonable differences, especially if we consider the size of the credible intervals. However, the fact that the differences are not large it is most likely due to the large size of the sample. But, it is possible to appreciate these discrepancies which show that different models for the bulk data impacts on the threshold value, as expected.

%--- DISCUSSION -----------------------------------------------
\section{Discussion}\label{sc_disc}
There are many processes which present heavy-tailed behaviour and, in these cases, it is not always advisable to represent the whole data by means of a parametric distribution, such as the Student-$t$, or by a mixture model where the components are densities of the same family, such as a mixture of gamma densities \citep{Ventur:2008,Castillo:2012} or normal densities.

One way to address the above problem is to consider the asymptotic result of Pickand's theorem, for which the tail of a distribution, above a certain threshold, can be represented by a GPD. However, this method raises another problem, that is the determination of an appropriate threshold. In a Bayesian set up, the idea is to represent the uncertainty about the threshold by a prior distribution.

In this paper, we present a way of defining prior distributions for the threshold of a GPD which have as support the set of observed order statistics. We propose two different methods to determine the prior: one is intuitive in the sense that every order statistic has, \textit{a priori}, the same probability of being the true threshold value. The second method takes into consideration the loss that we would incur if a given order statistic were removed from the set of possible values for the threshold, and it is true value. Through simulation and real data analysis, we have shown that the two priors have very similar performances, when compared on the basis of the frequentist properties of the respective posterior they yield and the estimates they generate. Given that the idea behind these priors is to represent a condition of minimal prior information, the fact that both priors converge to similar results is comforting. Nonetheless, it is still possible to find reasons to prefer one over the another.
The uniform prior has the undoubted advantage of being easier to code and, although minimally, it allows for faster simulations. However, one has to be careful in assuming that uniformity represents no prior information \citep{BerSmi:1994}. Thus, although the results obtained by applying the two discrete priors for the threshold are similar, we believe that the prior based on losses has to be preferred on the basis of the following considerations. First, it has a ``meaning''. The mass assigned to each order statistic derives from a sound consideration of the \emph{worth} that each one of them has in representing the potential threshold for the GPD. On the other hand, and this connects to the second reason, by assigning a uniform prior to $x^{(k)}$ one assumes that each order statistics has the same chance of being the threshold. Apart that it could be interpreted as an informative assumption, it conflicts with the idea behind the GPD, for which the threshold has to be sufficiently large to avoid model bias, as discussed in Section \ref{sc_intro}; and this is not compatible with a uniform prior where lower order statistics have an \emph{a priori} probability of being the threshold equal to the one of the upper order statistics.. In conclusion, when one aims for objectivity, in applied statistics problems a noninformative prior has to be based on solid motivations, not only on performance.
One exception to the above argument is the case of $\xi<0$ (which however is not in the scope of this paper). In fact, given that the parameter space for the threshold depends on the values of $\xi$ and $\sigma$, the prior based on losses would not be defined as the Kullback--Leibler divergence between different GPD is infinite. Thus, in the case we would model a light-tail of a distribution by a GPD, the choice of the uniform prior would be the only choice between the two proposed discrete prior distributions.

As a final remark, we would like to highlight that, although the focus of the paper has been on observations that can take positive values only, the overall approach can be easily extended to quantities over the whole real line. For example, if we consider logarithmic returns of some financial index (or price), the part of the data below the threshold could be modelled by a finite mixture of normal densities. Similarly, if we were interested in analysing the negative returns (which is a common practice in risk management, for example), the prior would be defined over the negative order statistics, and the bulk data would be represented by the observations above the threshold. Another possible development of the model and the inferential procedure would be represented by the case where both tails of the distribution would be of interest, and therefore represented by one separate, but not necessarily independent, GPD each. In all the above cases, the support of the prior has to be defined so to reflect the nature of the problem.

\vspace{1cm}
\noindent
\textbf{Acknowledgements}

\noindent
The author is very thankful to Professor Philip Brown for his valuable feedback and comments during the drafting of the paper.

\end{document}